\documentclass[pdflatex,sn-mathphys-num]{sn-jnl}

\usepackage{graphicx}%
\usepackage{multirow}%
\usepackage{amsmath,amssymb,amsfonts}%
\usepackage{amsthm}%
\usepackage{mathrsfs}%
\usepackage[title]{appendix}%
\usepackage{xcolor}%
\usepackage{textcomp}%
\usepackage{manyfoot}%
\usepackage{booktabs}%
\usepackage{algorithm}%
\usepackage{algorithmicx}%
\usepackage{algpseudocode}%
\usepackage{listings}%
\usepackage{multicol}
\usepackage{orcidlink}
\usepackage{lineno}

\newcommand{\arcdeg}{\mbox{$^\circ$}}%

\newcommand{\isro}{{\it ISRO }}
\newcommand{\suit}{{\it SUIT }}
\newcommand{\js}[1]{{#1}}
\graphicspath{{./}{pics/}}

\begin{document}

\title[Article Title]{Science Filter Characterization of the Solar Ultraviolet Imaging Telescope (SUIT) on board Aditya-L1.}


\author*[1, 2]{\fnm{Janmejoy}~\sur{Sarkar}~\orcidlink{0000-0002-8560-318X}}\email{janmejoy.sarkar@iucaa.in}
\author[1]{\fnm{Rushikesh}~\sur{Deogaonkar}~\orcidlink{0009-0000-2781-9276}}\email{rushikesh@iucaa.in}
\author[1]{\fnm{Ravi}~\sur{Kesharwani}~\orcidlink{0009-0002-2528-5738}}\email{ravik@iucaa.in}
\author*[3]{\fnm{Sreejith}~\sur{Padinhatteeri}~\orcidlink{0000-0002-7276-4670}}\email{sreejith.p@manipal.edu}
\author[1, 4]{\fnm{A.~N.}~\sur{Ramaprakash}~\orcidlink{0000-0001-5707-4965}}\email{anr@iucaa.in}
\author[1, 4]{\fnm{Durgesh}~\sur{Tripathi}~\orcidlink{0000-0003-1689-6254}}\email{durgesh@iucaa.in}
\author[1]{\fnm{Soumya}~\sur{Roy}~\orcidlink{0000-0003-2215-7810}}\email{soumyaroy@iucaa.in}

\author[2]{\fnm{Gazi A.} \sur{Ahmed}\orcidlink{0000-0002-0631-4831}}
\author[5]{\fnm{Rwitika} \sur{Chatterjee}}
\author[1, 4]{\fnm{Avyarthana} \sur{Ghosh} \orcidlink{0000-0002-7184-8004}}
\author[5]{\fnm{Sankarasubramanian} \sur{K.}}
\author[1]{\fnm{Aafaque} \sur{Khan}}
\author[1]{\fnm{Nidhi} \sur{Mehandiratta}}
\author[5]{\fnm{Netra} \sur{Pillai}}
\author[5]{\fnm{Swapnil} \sur{Singh} \orcidlink{0000-0002-3152-0477}}

\affil[1]{\orgname{The Inter-University Centre for Astronomy and Astrophysics}, \orgaddress{\street{Ganeshkind}, \city{Pune}, \postcode{411007}, \state{Maharashtra}, \country{India}}}
\affil[2]{\orgdiv{Department of Physics}, \orgname{Tezpur University}, \orgaddress{\street{Napaam}, \city{Tezpur}, \postcode{784028}, \state{Assam}, \country{India}}}
\affil*[3]{\orgdiv{Manipal Centre for Natural Sciences}, \orgname{Manipal Academy of Higher Education}, \orgaddress{\city{Manipal}, \postcode{576104}, \state{Karnataka}, \country{India}}}
\affil[4]{\orgdiv{Center of Excellence in Space Sciences India (CESSI)}, \orgname{Indian Institute of Science Education and Research}, \orgaddress{\city{Mohanpur}, \postcode{741246}, \state{West Bengal}, \country{India}}}
\affil[5]{\orgdiv{Space Astronomy Group}, \orgname{U. R. Rao Satellite Centre}, \orgaddress{\street{Vimanapura}, \city{Bengaluru}, \postcode{560017}, \state{Bengaluru}, \country{India}}}

\abstract{The Solar Ultraviolet Imaging Telescope (\suit) on board the Aditya-L1 mission is designed to observe the Sun across 200{--}400~nm wavelength. The telescope used 16 dichroic filters tuned at specific wavelengths in various combinations to achieve its science goals. For accurate measurements and interpretation, it is important to characterize these filters for spectral variations as a function of spatial location and tilt angle. Moreover, we also measured out-of-band and in-band transmission characteristics with respect to the inband transmissions. In this paper, we present the experimental setup, test methodology, and the analyzed results. Our findings reveal that the transmission properties of all filters meet the expected performance for spatial variation of transmission and the transmission band at a specific tilt angle. The out-of-band transmission for all filters is below 1\% with respect to in-band, except for filters BB01 and NB01. These results confirm the capabilities of \suit to effectively capture critical solar features in the anticipated layer of the solar atmosphere.}


\maketitle

\section{Introduction}\label{sec:intro}
The Solar Ultraviolet Imaging Telescope \cite{suit17, suitghosh2016} \suit on board Indian Space Research Organization's (ISRO's) Aditya-L1 \cite{aditya, adityal1} makes 24 $\times$ 7 full-disk observations of the Sun in eleven spectral bands (Table \ref{tab:science_filters}) within the near ultraviolet band of 200 - 400 nm \cite{suitghosh2016} from the Sun-Earth Lagrange 1 point. To date, a limited number of imaging facilities have been built to operate in this wavelength band.
\textit{The Interface Region Imaging Spectrometer (IRIS)} \cite{iris} and the \textit{Ultraviolet Spectrometer and Polarimeter (UVSP)} \cite{uvsp} operate in this band, but have a limited spectral and spatial coverage.
The balloon-borne \textit{Sunrise} mission too covered this complete wavelength range. The \textit{Sunrise Filter Imager (SuFI)} facilitates observation at 214 nm with 10 nm bandwidth, 300 nm with 5 nm bandwidth, 388 nm with 0.8 nm bandwidth, and 396.8 nm (core of Ca II h) at 0.18 nm bandwidth \cite{Solanki_2010}, which are similar to some of the observation bands of \suit as mentioned in Table \ref{tab:science_filters}. However, the observations from \textit{Sunrise} were for a short period of time, for a small field of view, albeit with high resolution. 

\begin{figure}
    \centering
    \includegraphics[width=0.7\linewidth]{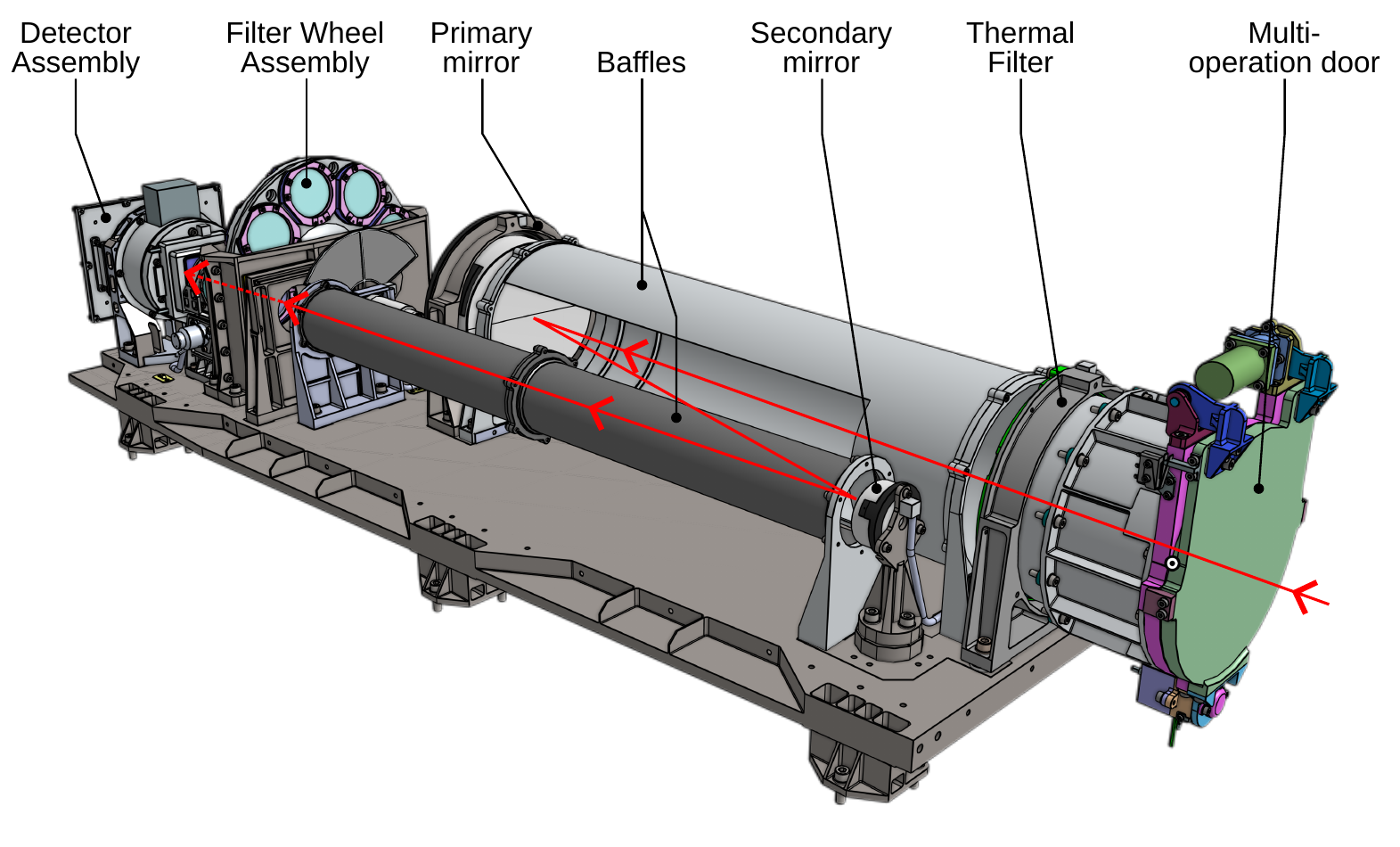}
    \caption{Optical path of sunlight entering the \suit payload for imaging. Incident light is attenuated by the thermal filter and gets reflected by the primary and secondary mirrors. The light passes through the science filters, followed by a field correction lens, which is focussed on the CCD.}
    \label{fig:ray_path}
\end{figure}

\suit employs an off-axis Ritchey Chr\'{e}tien optical design and provides a field of view of $\pm 1.5 R_\odot$ \js{from the field center}, with a plate scale of 0.7"/pixel. This allows full disk observations at high cadence, currently unavailable with any other instrument operating in this wavelength band. \suit has a thermal filter \cite{thermalfilter_padinhatteeri} at its entrance aperture to filter visible and IR while passing a fraction of the incident UV. This helps to keep the temperature of the optical cavity under control and prevents the saturation of the CCD.
This filter gives a transmission of 0.1-0.45 \% within the 200-400 nm band while giving a transmission of 0.2\% in visible light \cite{varma2023} \js{and not more than 0.4\% between 800-1200 nm \cite{thermalfilter_padinhatteeri}}. After entering, this light is reflected off the primary and secondary mirrors, before being incident on the science filters mounted on the filter wheel assembly. The filters help in transmitting the desired band of light which then passes through a position adjustable field corrector lens, before reaching the CCD. \js{A schematic diagram of the ray path is illustrated in Figure \ref{fig:ray_path}.}

The filter wheel assembly has two stacked filter wheels holding eight filters each. Combinations of these sixteen filters generate the eleven desired bandpasses for SUIT \cite{suit17}. These science filters are manufactured by Materion Corp. as per specifications defined by the \suit team. Each filter has an outer diameter of 53 mm and a clear aperture of 49 mm. Two coated UV-fused silica components are cemented into one black anodized aluminum with a vented air gap in between. This gives the filter glass a thickness of 3.5 mm while the ring is 5 mm thick.	Each glass substrate has a transmission wavefront of $< \lambda/4$ per inch, surface roughness of $< 2nm$ RMS, and a wedge uncertainty of $<\pm 10"$ measured before coating. The spectral shift of the filters with temperature is $\sim 2~pm/^\circ C$. The corresponding spectral fluctuation is negligible with the onboard operational temperature range being \js{$(20 \pm 1)^\circ C$}. The filters can be safely stored within a temperature range of  $0^\circ - 80^\circ$  at a relative humidity of $<50\%$ in a Class 100-like clean atmosphere.


In this paper, we discuss the filter qualification tests in \S \ref{sec:qual_test} and the experimental setup in \S \ref{sec:ExptSetup}. We discuss the test procedure and results for spectral and spatial Characterization of the filters in \S \ref{sec:Spatial}, out-of-band characterization in \S \ref{sec:oob}, and Transmission variation with tilt angle in \S \ref{sec:tilt}. The methodology used for mounting the filters in the payload is discussed in \S \ref{sec:mounting}. The paper concludes with the summarized performance and acceptability of the filter performance in \S \ref{sec:conclusion}. The optical throughput and spectral transmission of the complete payload in eleven bandpasses are discussed separately in \cite{photometric_sarkar}.

\begin{table*}
\label{tab:science_filters}
\centering
\begin{tabular}{c c c c c}
\toprule
\textbf{Science}  &	\textbf{Combination} &	\textbf{Central} & \textbf{Bandpass} &\textbf{Science} \\
\textbf{Filter}	&	\textbf{Filter}     &	\textbf{Wavelength  (nm)}	&		\textbf{(nm)	}	   	&\textbf{target}		\\
\midrule
NB01    & BB01 	& 214.0 	& 11.0 	& Continuum\\
NB02 	& BP02	& 276.6		& 0.4 	& Mg~\rm{II}~k blue wing \\
NB03 	& BP02	& 279.6 	& 0.4 	& Mg~\rm{II}~k\\
NB04 	& BP02	& 280.3		& 0.4 	& Mg~\rm{II}~h\\
NB05	& BP02	& 283.2		& 0.4 	& Mg~\rm{II}~h red wing\\
NB06 	& BP03	& 300.0 	&1.0 	& Continuum\\
NB07 	& BP03	& 388.0		&1.0 	& CN Band\\
NB08	& NB08	& 396.85 	& 0.1 	& Ca~\rm{II}~h\\
BB01 	& BB01	& 220.0		& 40.0	& Herzberg Continuum \\
BB02 	& BP04	& 277.0 	& 58.0  & Hartley Band\\
BB03 	& BP04	& 340.0		& 40.0  & Huggins Band\\
\botrule
\end{tabular}
\caption{List of \suit bandpasses and the facilitating filters with the corresponding central wavelengths, transmission bandwidth (Full Width at Half of Maximum for Narrowband \js{(NB)} filters and 99\% transmission bandwidth for Broadband \js{(BB)} and Bandpass \js{(BP)} filters), and the observation interest.} 
\end{table*}

\section{Science Filter Qualification Tests}\label{sec:qual_test}
Environmental tests are performed to qualify the filters and the filter coatings to withstand the space environment the filters would withstand. Most of the qualification tests were performed by the vendor on all the delivered filters, except for a proton radiation test, which was performed by the payload team on representative filter witness coupons.
The filters are subjected to a thermovacuum test at a pressure of $10^{-5}$ mbar. \js{The mean operating temperature inside the optical cavity of \suit is $\sim (20 \pm 1) ^\circ C$.} So, the filters are exposed to 5 thermal cycles between \js{$10^\circ C - 30^\circ C$} ramped at $2^\circ$C/min with a dwell time of 2 hours \js{during environmental tests. }
Humidity test was performed by exposing the filter to 90\% relative humidity at a temperature of $40^\circ C \pm 2 ^\circ C$ for 24 hours.
Coating durability \js{tests} for adhesion and abrasion of the coating were performed as per MIL-C-675C standards. The test parameters are summarized in Table \ref{tab:qual_test}. All the mounted filters comply with these standards and have survived these environmental tests. 

\js{A proton} radiation test was performed by the payload team at the Tata Institute of Fundamental Research, Mumbai, on filter witness coupons provided by the vendor.  The displacement damage equivalent fluence was $5.6 \times 10^{10}~protons/cm^2$ for 10 MeV \js{protons}. The filter coupons were also exposed to a direct ionization exposure of $\sim 27$ krad at a rate of 168 rad/sec for the gamma testing. The test results are summarized in Table \ref{tab:proton_test}. The extent of variation in total transmission, central transmission wavelength, and bandwidth are expected to make an \js{negligible} variation in transmission performance. \js{A variation of 2.28\% in peak transmission is seen for NB06 after the proton irradiance test. This is within the tolerable limits of ($30 \pm 5$)\%. To monitor such variations in peak transmission levels over time, \suit shall periodically observe Sirius to perform recalibration.}

\begin{table}
    \centering
    \begin{tabular}{c|c}
        \toprule
        \textbf{Test} & \textbf{Test Specifications} \\
        \midrule
        \multicolumn{2}{c} {\textit{Thermo-vacuum Test}}\\
        \midrule
        No. of Cycles & 5\\
        Temperature Range & $10^\circ C-30 ^\circ C$\\
        Dwell time & 2 hours\\
        Rate of temperature change & $\sim 2^\circ C/min$\\
        Vacuum Level & $< 10^{-5}mbar$\\
        \midrule
        \multicolumn{2}{c} {\textit{Humidity Test}}\\
        \midrule
        Relative Humidity & 95\%\\
        Temperature & $40^\circ C \pm 2^\circ C$\\
        Exposure duration & 24 hrs\\
        \midrule
        \multicolumn{2}{c} {\textit{Coating Durability Test}}\\
        \midrule
        Adhesion and Abrasion tests & As per MIL-C-675C standard\\
        \midrule
        \multicolumn{2}{c} {\textit{Radiation Test}}\\
        \midrule
        Total ionization dose hardness & 27 krad at 168 rad/s\\
        Displacement Damage Equivalent & \\
        Fluence (10 MeV Proton) & $5\times 10^{10} protons/cm^2$\\
        \bottomrule
    \end{tabular}
    \caption{Environmental tests performed on \suit science filters. The first column mentions the test performed, and the second column presents the test conditions.}
    \label{tab:qual_test}
\end{table}

\begin{table}
    \centering
    \begin{tabular}{c|c|c|c|c|c|c|c|c|c}
        \toprule
        \textbf{Filter} & \multicolumn{3}{|c|}{\textbf{Peak Wavelength (nm)}} & \multicolumn{3}{|c|} {\textbf{FWHM (nm)}} & \multicolumn{3}{|c}{\textbf{Peak Transmission}} \% \\
        \midrule
        & \textit{Design} & \textit{Pre} & \textit{Post} & \textit{Design} & \textit{Pre} & \textit{Post} & \textit{Design} & \textit{Pre} & \textit{Post}\\
        \midrule
        NB03 & $279.6 \pm 0.1$ & $279.49$ & $279.54$ & $0.4 \pm 0.1$ & $0.46$ & $0.44$ & $26\pm5$ & $15.21$ & $15.68$\\
        NB06 &\js{ $300.0\pm0.15$ }& $299.83$ & $299.83$ &\js{ $1.0\pm0.2$ }& $1.15$ & $1.11$ & $30\pm5$ & $30.99$ & $28.71$\\
        NB07 &\js{ $388.0\pm0.15$ }& $387.57$ & $387.63$ &\js{ $1.0\pm0.2$ }& $1.06$ & $1.04$ & $30\pm5$ & $21.24$ &\js{ $21.20$ }\\
        \bottomrule
    \end{tabular}
    \caption{Proton irradiance test results for \suit science filter witness coupons. The peak wavelength, FWHM of the transmission curve, and the peak percentage of transmission are measured before and after tests and compared with the design values.}
    \label{tab:proton_test}
\end{table}

\section{Experimental Setup} \label{sec:ExptSetup}
The \suit payload has a collection of 11 science bandpasses aiming to observe different layers of the solar atmosphere with its eight narrowband filter combinations and study the effect of solar irradiance on Earth's climate with three broadband filter combinations. These combinations are facilitated by sixteen dichroic science filters mounted on two 8-position filter wheels. 

\begin{figure}
    \centering
    \includegraphics[width=1\linewidth]{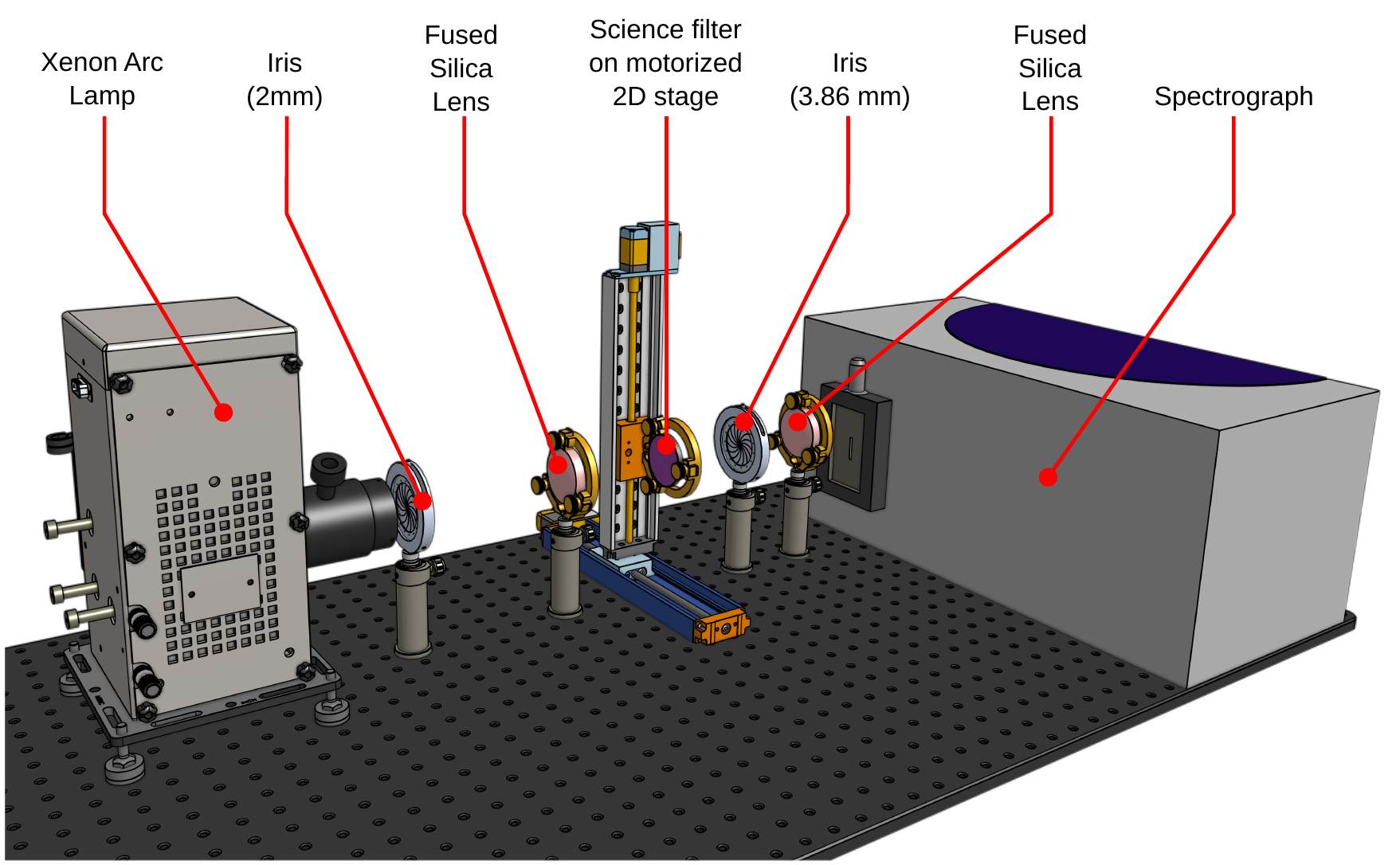}
    \caption{Schematic representation of filter characterization setup. Light from the Xe Lamp passes through an iris before being collimated by a lens. The collimated light passes through the filter. An iris limits the transmitted light to a diameter of 3.86 mm. The light is focussed into the slit of the spectroscope with a focusing lens. }
    \label{fig:filterschematic}
\end{figure}

The filters are spectroscopically characterized before mounting on the filter wheels at the Class 100 clean room facility at \isro Satellite Integration and Testing Establishment in Bengaluru, India. Three aspects of the filters are measured here- The variation of transmission across five different spatial locations on the filter, the out-of-band transmission on the blue and red wings of the filter transmission curve as a percentage of the inband transmission, and the variation of transmission spectrum with a change in the tilt angle of the filter with respect to the incident light.
The operating wavelength range of the science filters makes it susceptible to molecular and particulate contamination. Therefore, the experimental setup is enclosed and purged with 99.999\% pure gaseous nitrogen to maintain a positive pressure, preventing any contaminants from entering the test setup. A high-resolution Andor Shamrock 500i imaging spectrometer is used to perform the spectroscopic tests, with an appropriate optical setup and a stable Xenon Arc Lamp for illumination of the target. An iris of 2 mm aperture is placed at the face of the lamp to restrict stray light from entering the optical setup. A fused silica lens with high UV transmission collimates the light from the iris. This collimated beam falls on the science filter, mounted on a motorized translation stage for spatial transmission variation measurements or a calibrated rotation stage for measuring transmission at various tilt angles. Light transmitted from the filter is stopped down by an iris with an aperture of 3.86 mm. This is the spot size at the filter plane of SUIT for light coming from one resolution element of \suit (0.7") from the Sun. This light is focused onto the slit of the spectrometer with a fused silica lens. Figure \ref{fig:filterschematic} shows a schematic diagram of the experimental setup.

Measurements are taken with two gratings in the spectrometer- 1200 lines per mm blazed at 500 nm, and 2400 lines per mm holographic grating blazed at 220 nm.
These gratings provide a spectral coverage of 40 nm and 20 nm, with spectral resolutions of 0.02 nm and 0.01 nm, respectively. The choice of grating is based on the transmission characteristics of the filter and the spectrum to be recorded. Before the measurements, wavelength calibration is performed separately with an Ocean Optics HG-2 (Mercury-Argon) wavelength calibration lamp for both gratings.

Broadly, four types of images are recorded to measure the percentage of filter transmission as depicted in Equation \ref{eqn:tx}. The spectra are recorded with ($T_f$) and without ($T_s$) the filter in the beam path with exposure times $ E_f $ and $ E_s $, respectively. The spectroscope slit is covered with an opaque obstruction, and the background signal is recorded at the same exposure times, which are denoted by $B_f$ and $B_s$, respectively. The background spectra are removed from the corresponding transmission spectra, which are then normalized with exposure time to get the \js{transmission:}

    \begin{align} 
        Tx&= \dfrac{(T_f-B_f)/ E_f}{(T_s- B_s)/ E_s} \label{eqn:tx}\\
        \nonumber where,&\\
        \nonumber Tx&= Transmission\\
        \nonumber T_f&= Measured ~ spectrum ~ with ~ the ~ filter ~ in ~ the ~ beam ~ path\\
        \nonumber E_f&= Exposure ~ time ~ for ~ T_f\\
        \nonumber B_f&= Background ~ spectrum ~ with ~ exposure ~ time ~ E_f\\
        \nonumber T_s&= Measured ~ spectrum ~ without ~ the ~ filter ~ in ~ the ~ beam ~ path\\
        \nonumber E_s&= Exposure ~ time ~ for ~ T_s\\
        \nonumber B_s&= Background ~ spectrum ~ with ~ exposure ~ time ~ E_s\\
\end{align}

All the spectral measurements are made in air, while the filters are designed to operate in vacuum. As the filters are dichroic, the transmission wavelength increases when the operation environment changes from air to vacuum by \js{order of $10^{-1}$ nm \cite{mcloed_filt}}. Also, the transmission bandpass can change with operational \js{temperature by an order of $10^{-3}~nm/^\circ C$ \cite{mcloed_filt}}. \js{The values for these wavelength shifts were provided by the vendor and were implemented while calculating the transmissions.}
The operating temperature of the filters in space is $20^\circ C \pm 1^\circ C$, the same as the temperature at which the transmissions were measured in the lab. Therefore, the effective shift in wavelength due to operating temperature during the experiment is negligible.

\section{Spectral and Spatial Characterization} \label{sec:Spatial}
\begin{figure}
    \includegraphics[width = 0.45 \linewidth]{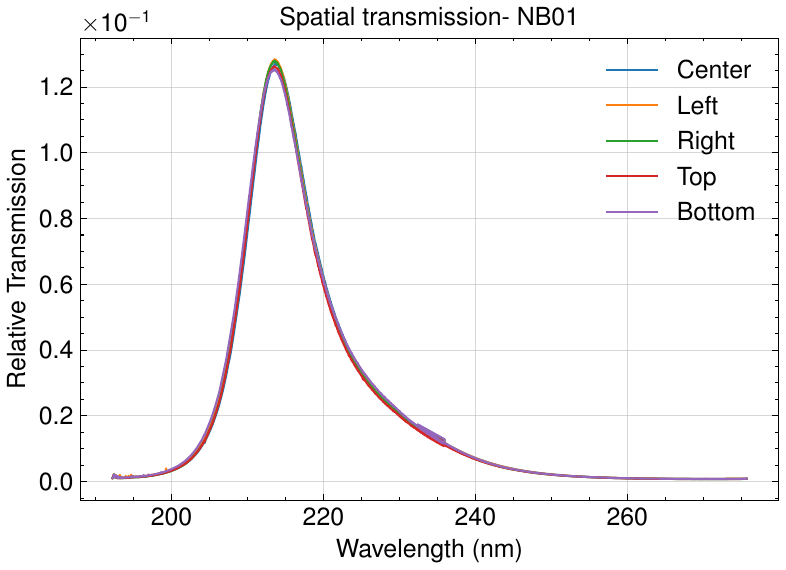}
    \includegraphics[width = 0.45 \linewidth]{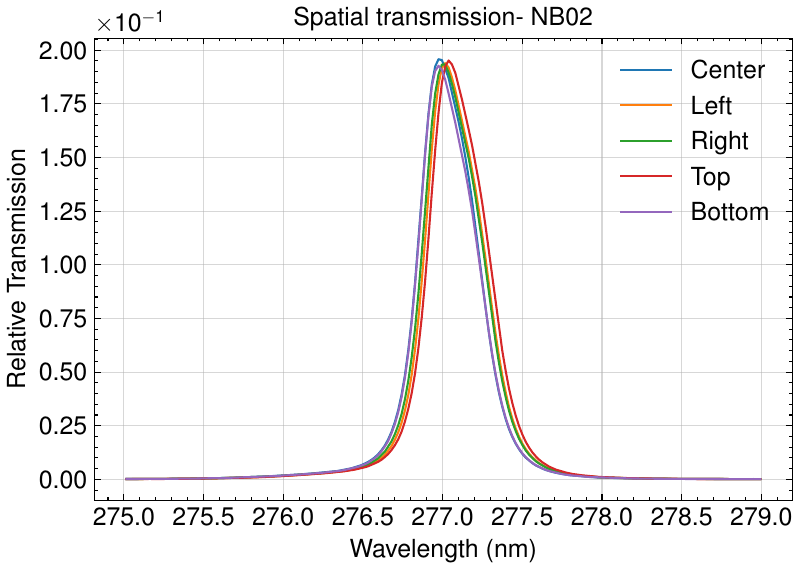}
    \includegraphics[width = 0.45 \linewidth]{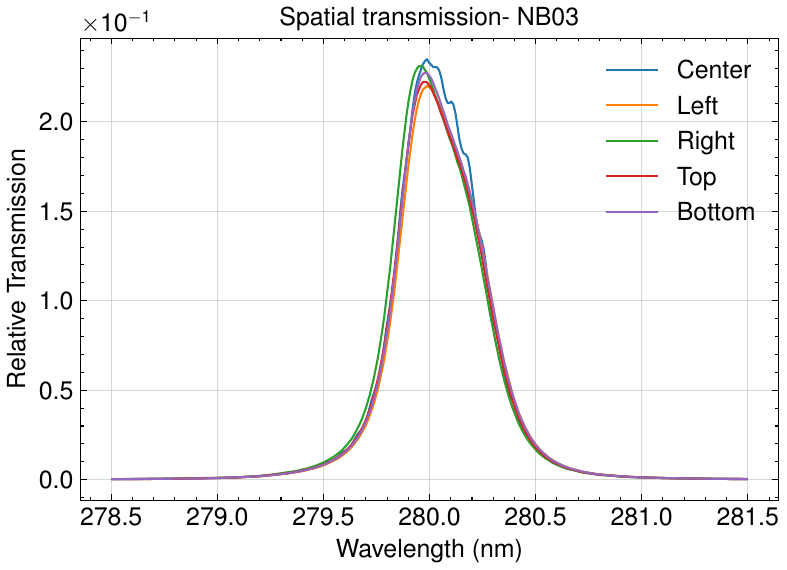}
    \includegraphics[width = 0.45 \linewidth]{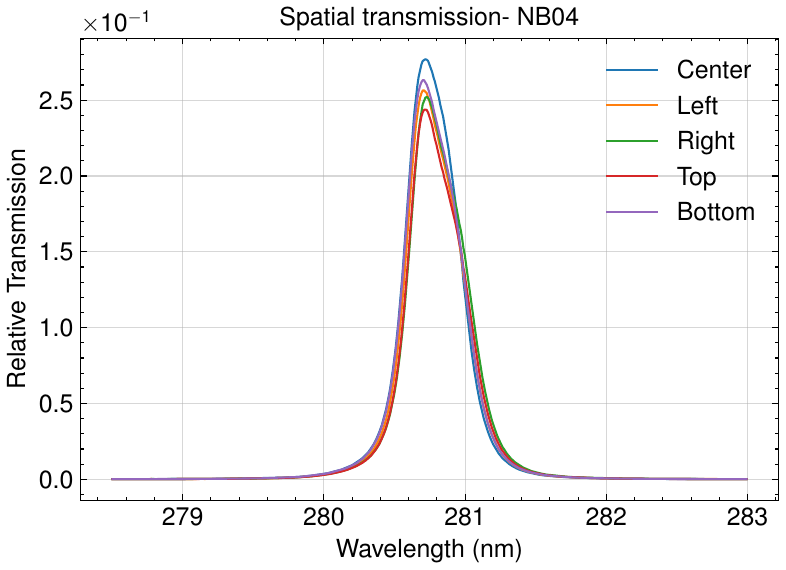}
    \includegraphics[width = 0.45 \linewidth]{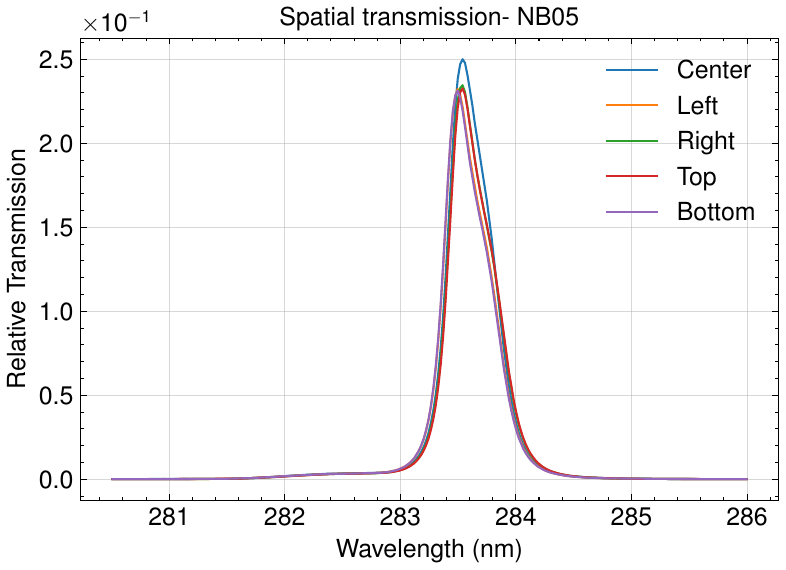}
    \includegraphics[width = 0.45 \linewidth]{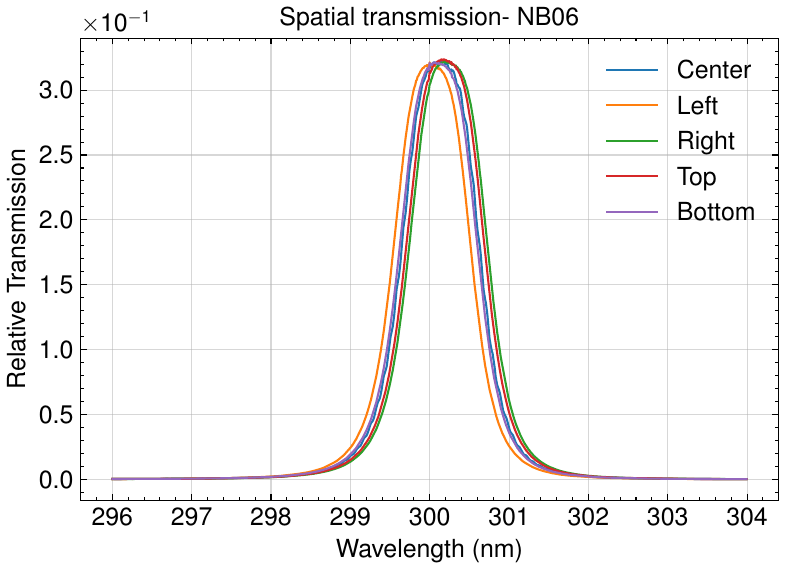}
    \includegraphics[width = 0.45 \linewidth]{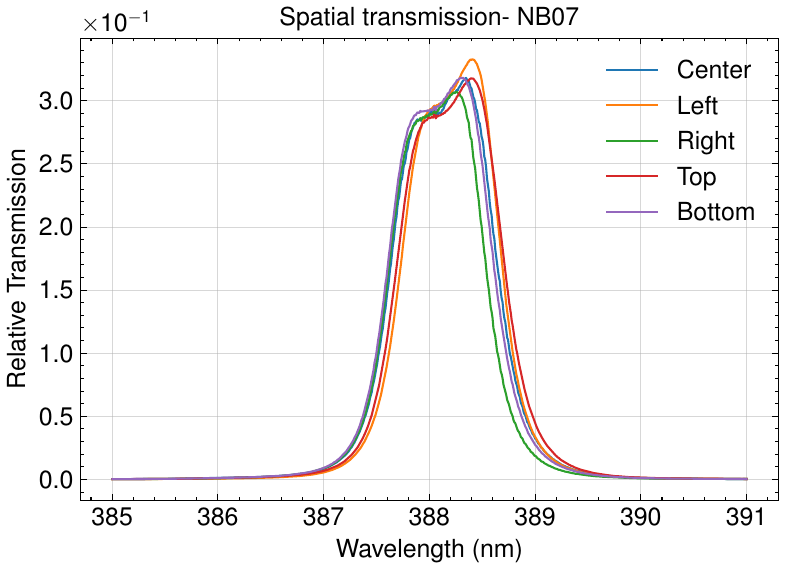}
    \includegraphics[width = 0.45 \linewidth]{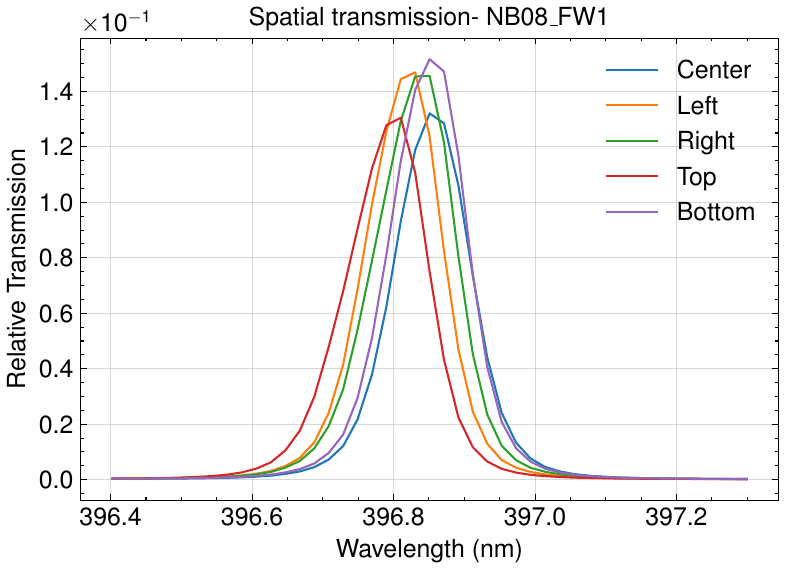}
    \includegraphics[width = 0.45 \linewidth]{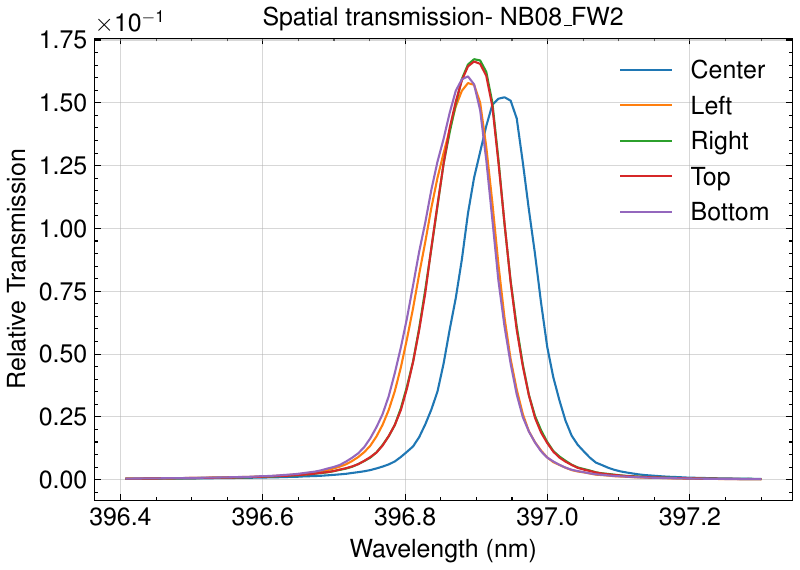}
    \caption{Spatial Variation of Transmission for \suit narrowband filters. The colors mark the transmission curves from various spatial locations of the filter. The \textit{y-axis} represents the relative transmission, while the \textit{x-axis} represents the wavelength in \textit{nm}.}
    \label{fig:spatial1}
\end{figure}
\begin{figure}
\centering
    \includegraphics[width = 0.45 \linewidth]{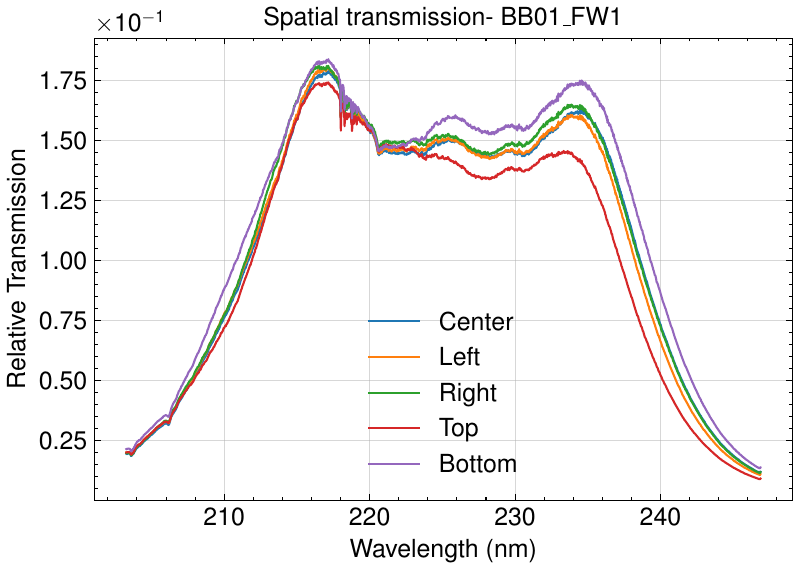}
    \includegraphics[width = 0.45 \linewidth]{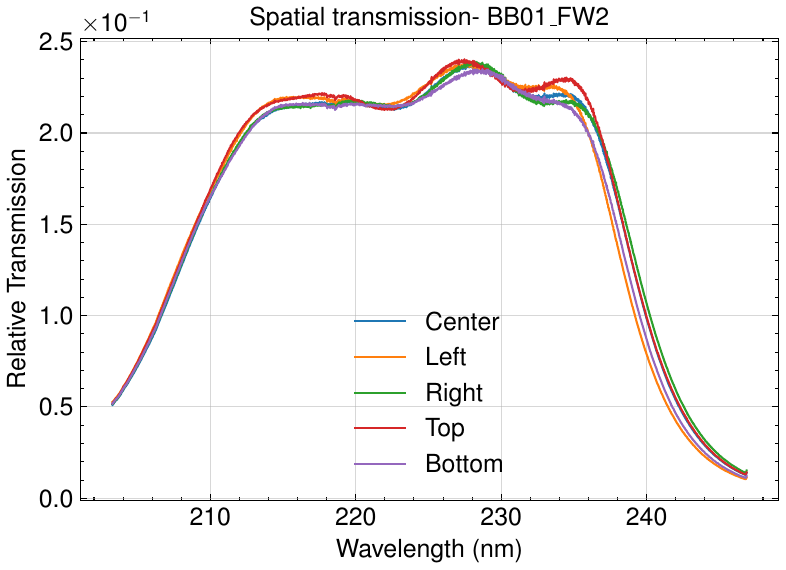}
    \includegraphics[width = 0.45 \linewidth]{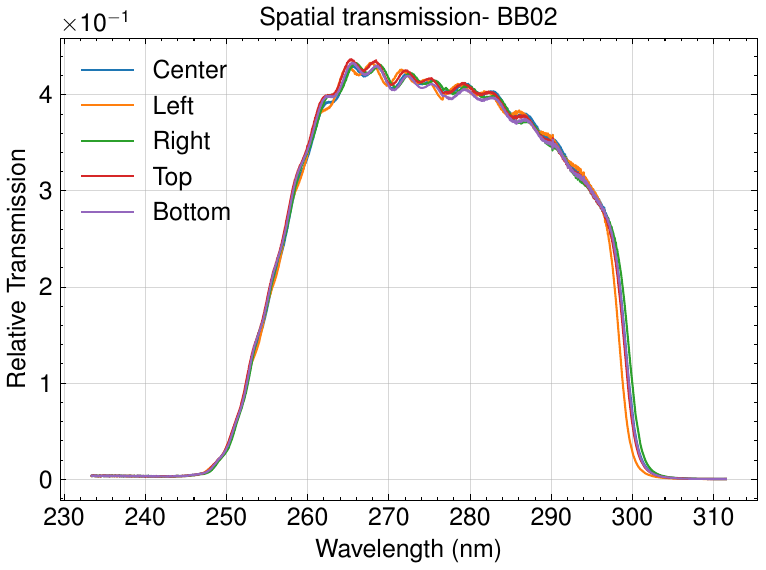}
    \includegraphics[width = 0.45 \linewidth]{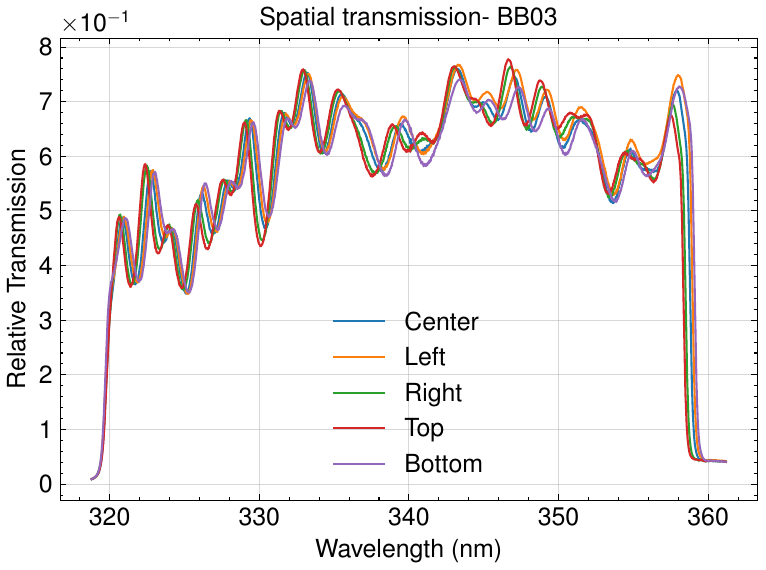}
    \includegraphics[width = 0.45 \linewidth]{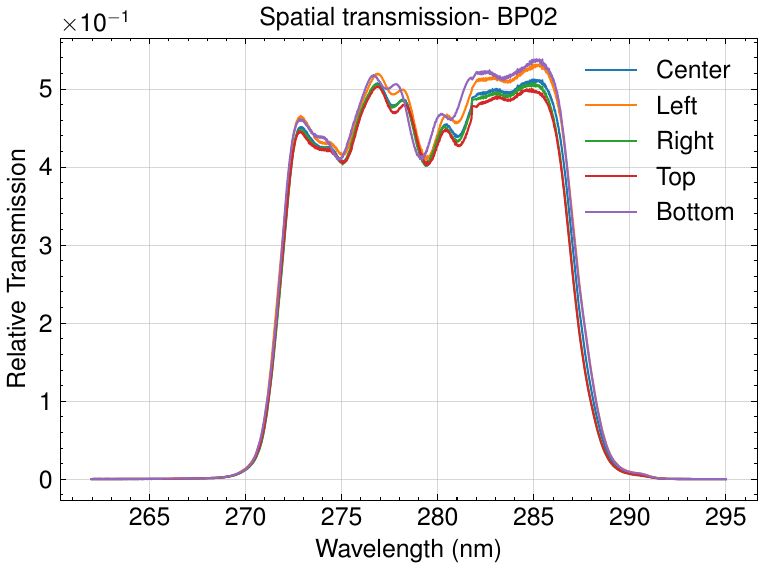}
    \includegraphics[width = 0.45 \linewidth]{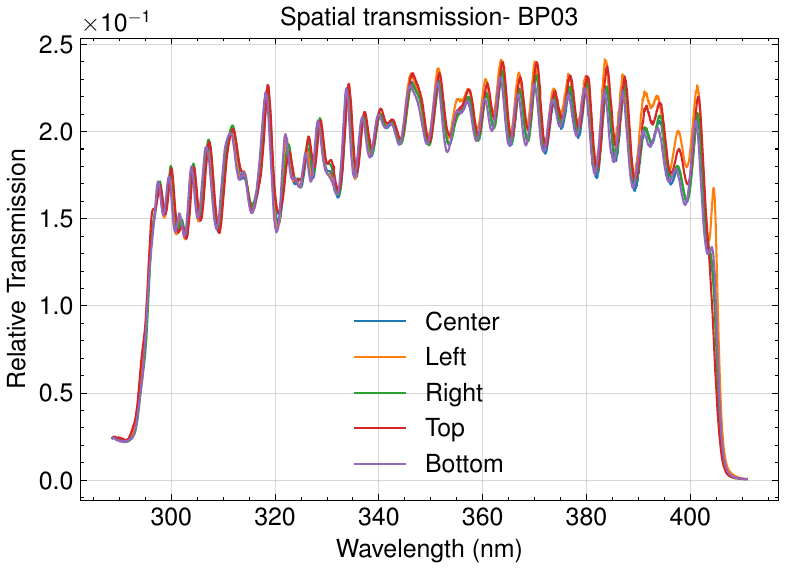}
    \includegraphics[width = 0.45 \linewidth]{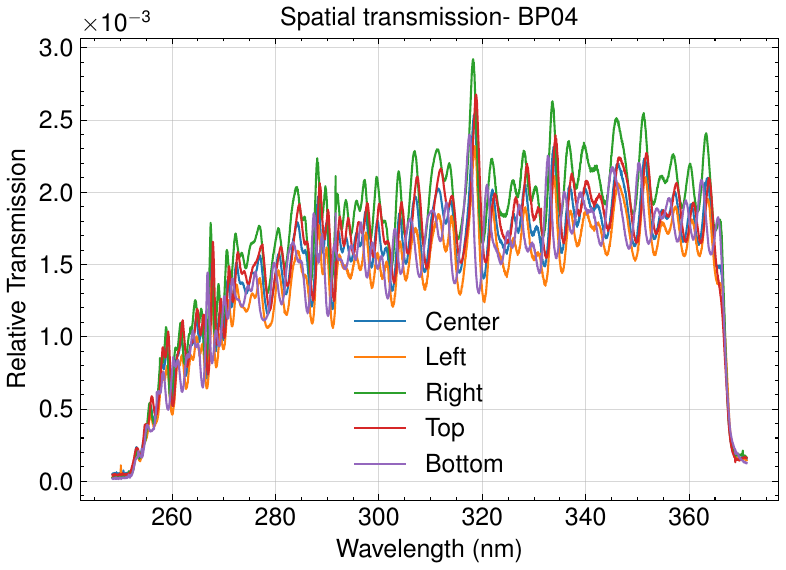}
    \caption{Spatial Variation of Transmission for \suit broadband and bandpass filters. The colors mark the transmission curves from various spatial locations of the filter. The \textit{y-axis} represents the relative transmission, while the \textit{x-axis} represents the wavelength in \textit{nm}.}
    \label{fig:spatial2}
\end{figure}
            
\subsection{Experiment}
A filter is expected to have the same transmission across its entire area. However, in real circumstances, the transmission may vary across the surface of the filter. This test is performed to characterize the variation of transmission of a filter at different spatial locations.

A filter is mounted on a motorized 2D stage, which allows precision horizontal and vertical movement automated with a computer. Each science filter has a radius of 26.5 mm. To spatially cover \js{$ > 65\%$} of the filter, readings are taken with the collimated light beam at the filter's center and at four locations 17.5 mm away from the center- towards the top, bottom, left, and right of the filter center. All the filters are mounted to the setup, \js{about a common marker on the rim of every filter with a specific marker on the filter mount,} to specifically identify the spatial locations. The imaging sequence records the transmission spectra, spectra of the light source, and the corresponding background frames. The percentage transmission at each spatial location is derived using Equation \ref{eqn:tx}, and the variation in transmission is monitored.

\subsection{Analysis and Results}
\begin{table}
\centering
\begin{tabular}{c c c c c c c}
    \toprule 
    \textbf{Filter} &  \multicolumn{2}{c}{\textbf{Peak Transmission}} &   \multicolumn{2}{c}{\textbf{Peak tx $\lambda$ (nm)}} & \multicolumn{2}{c}{\textbf{FWHM (nm)}}  \\ 
    \midrule
    & Mean & Deviation & Mean & Deviation & Mean & Deviation \\ 
    \midrule
    NB02 & 0.194 & $1.17e{-03}$ & 277.00 & $2.29e{-02}$ & 0.41 & $2.03e{-03}$ \\ 
    NB03 & 0.227 & $5.53e{-03}$ & 279.98 & $8.96e{-03}$ & 0.44 & $3.54e{-03}$ \\ 
    NB04 & 0.259 & $1.12e{-02}$ & 280.71 & $8.35e{-03}$ & 0.44 & $8.30e{-03}$\\ 
    NB05 & 0.236 & $7.01e{-03}$ & 283.52 & $2.04e{-02}$ & 0.43 & $4.32e{-03}$ \\
    NB06 & 0.322 & $8.55e{-04}$ & 300.11 & $6.81e{-02}$ & 1.04 & $6.25e{-03}$ \\ 
    NB07 & 0.319 & $8.23e{-03}$ & 388.34 & $5.42e{-02}$ & 1.01 & $3.85e{-02}$ \\ 
    NB08 \footnotemark[1]& 0.161 & $5.63e{-03}$ & 396.90 & $1.87e{-02}$ & 0.12 & $1.50e{-03}$ \\ 
    NB08 \footnotemark[2]& 0.143 & $8.64e{-03}$ & 396.84 & $2.09e{-02}$ & 0.13 & $2.90e{-03}$ \\ 
    \botrule 
\end{tabular} 
\footnotetext[1]{Mounted on filter wheel 1} 
\footnotetext[2]{Mounted on filter wheel 2}
\vspace{0.02\linewidth}
\caption{Deviation of filter transmission with spatial location. The columns indicate for all spatial locations of the NB filter in test, the mean and standard deviations of peak transmission [col 2 and 3],  the mean peak wavelength and its standard deviation [col 4 and 5], and the mean and standard deviations of the full width at half of the maximum transmission (FWHM) [col 6 and 7]. The relatively high deviation of the Peak transmission wavelength for NB08 with respect to the FWHM can be attributed to the extremely narrow bandpass ($\sim 0.1 nm$) of the filter, making it difficult to maintain the same transmission bandpass across the area of the filter.}
\label{table:deviation}
\end{table}	

The transmission variation with spatial location on the filter is evaluated by identifying the shift in central wavelength and the change in FWHM of the transmission profiles. The measured spectra are depicted in Figures \ref{fig:spatial1} and \ref{fig:spatial2}. The mean values and standard deviations for maximum transmission, peak transmission wavelength, and FWHM are calculated and \js{summarized} in Table \ref{table:deviation}.

It is noticed that the standard deviation of the wavelength for maximum transmission is of the order of $10^{-2}~nm$ for all the measured filters. The standard deviation in FWHM is of the order of $< 10^{-2}~nm$. These indicate that the spectral character of the filters does not vary drastically for different spatial locations. The variation in the percentage of peak transmission ranges in the order of $<1\%$ for different spatial locations in the narrowband filters. However, for NB08, we notice a relatively high deviation of the peak transmission wavelength with respect to the FWHM. This can be traced to the filter's extremely narrow bandpass ($\sim 0.1 nm$). Uniformly coating the filter substrate with the dichroic layers to achieve this wavelength band is particularly challenging, which can be attributed to the variation in peak transmission wavelength across the area of the filter.

The wavelength variation of transmission for different spatial locations of broadband and bandpass filters has little impact on the instrument throughput. Therefore, they are not \js{reported} in Table \ref{table:deviation}.

\section{Out of Band Characterization} \label{sec:oob}

\begin{figure}
    \centering
    \includegraphics[width = 0.45 \linewidth]{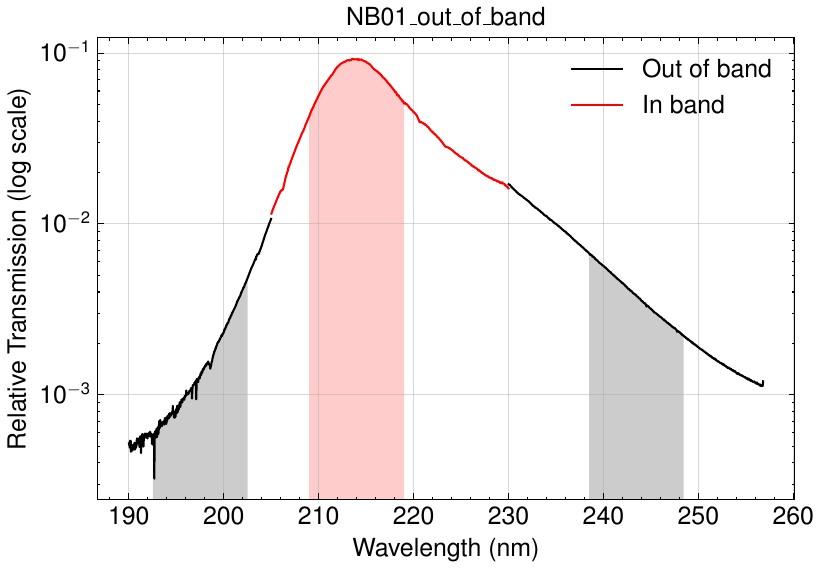}
    \includegraphics[width = 0.45 \linewidth]{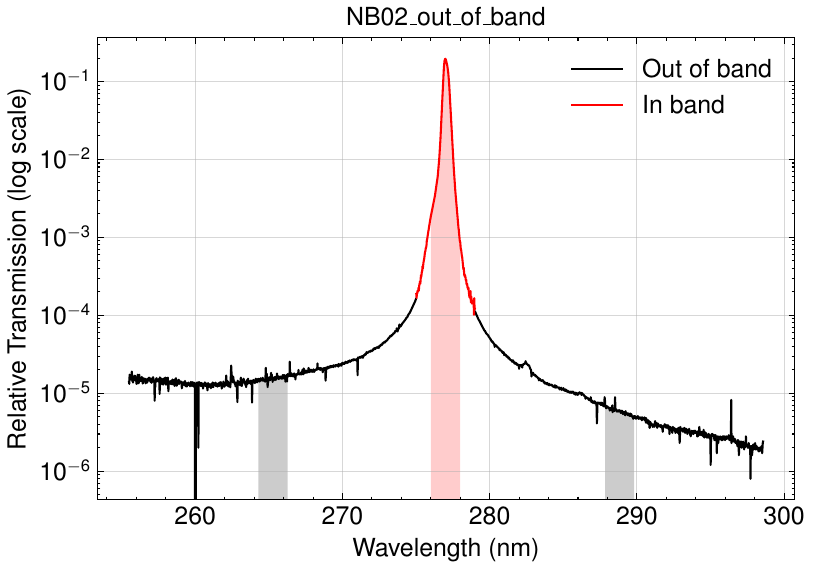}
    \includegraphics[width = 0.45 \linewidth]{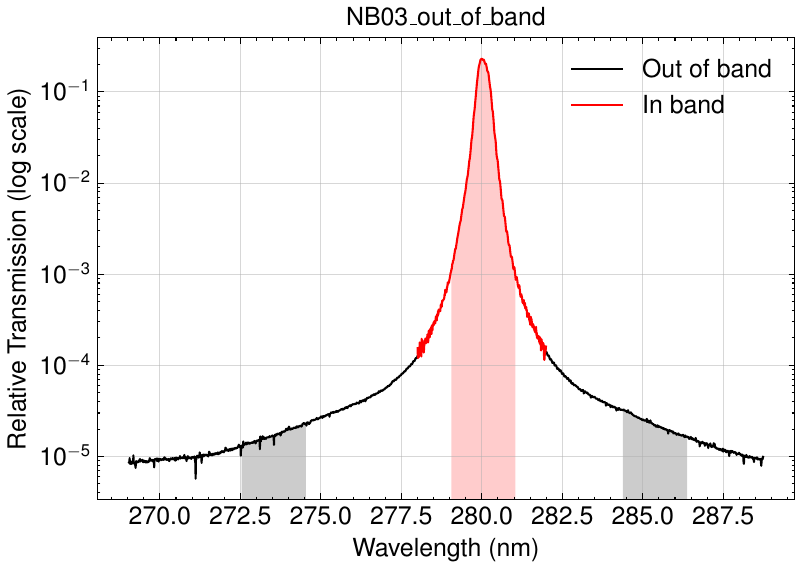}
    \includegraphics[width = 0.45 \linewidth]{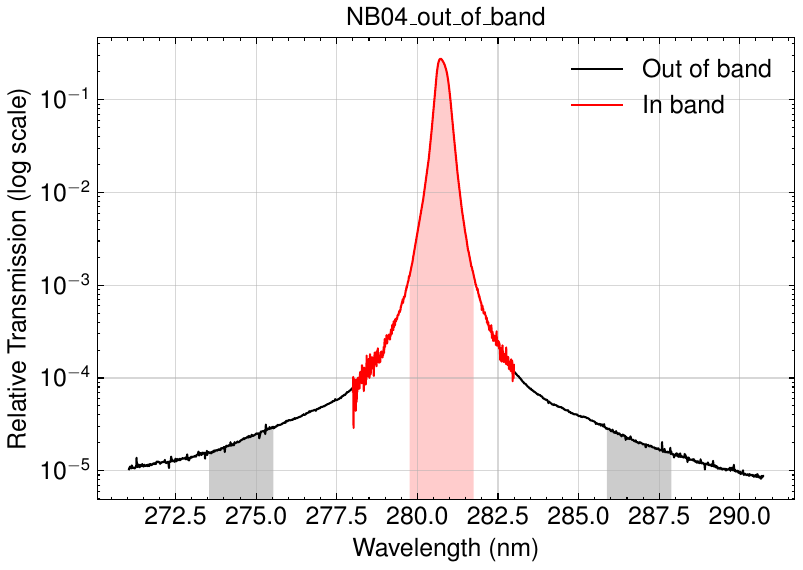}
    \includegraphics[width = 0.45 \linewidth]{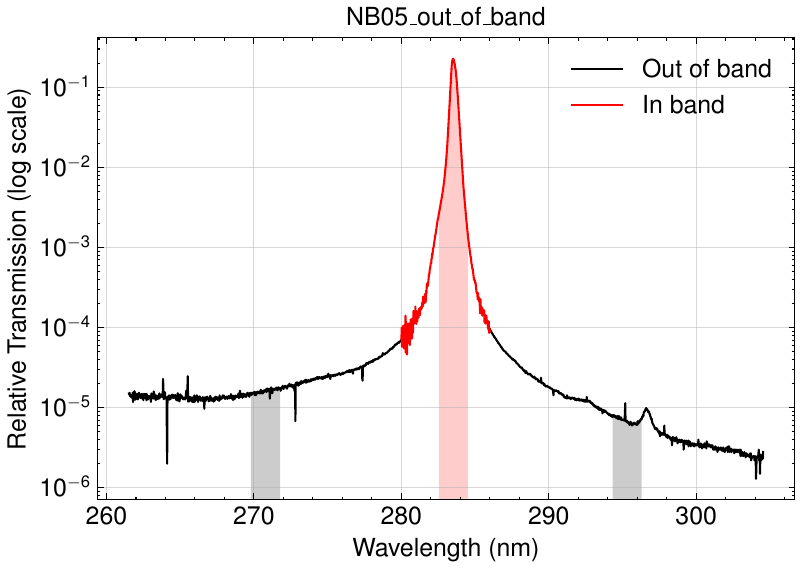}
    \includegraphics[width = 0.45 \linewidth]{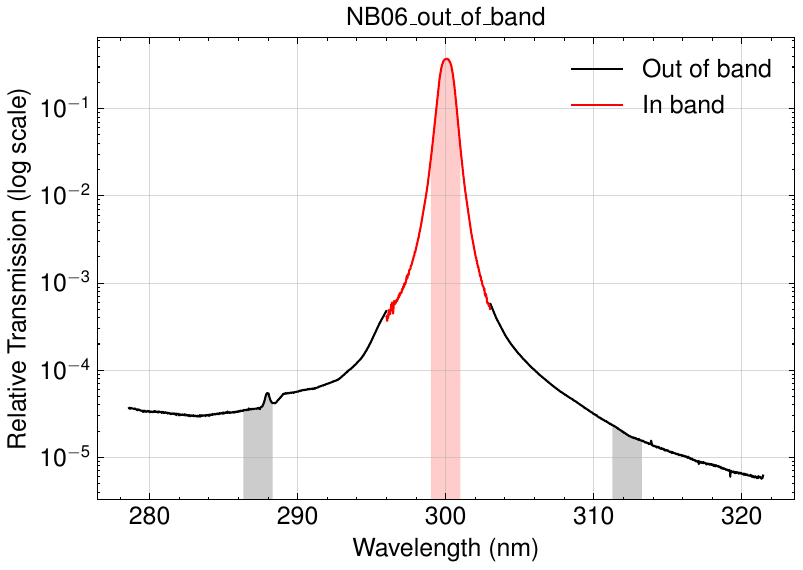}
    \includegraphics[width = 0.45 \linewidth]{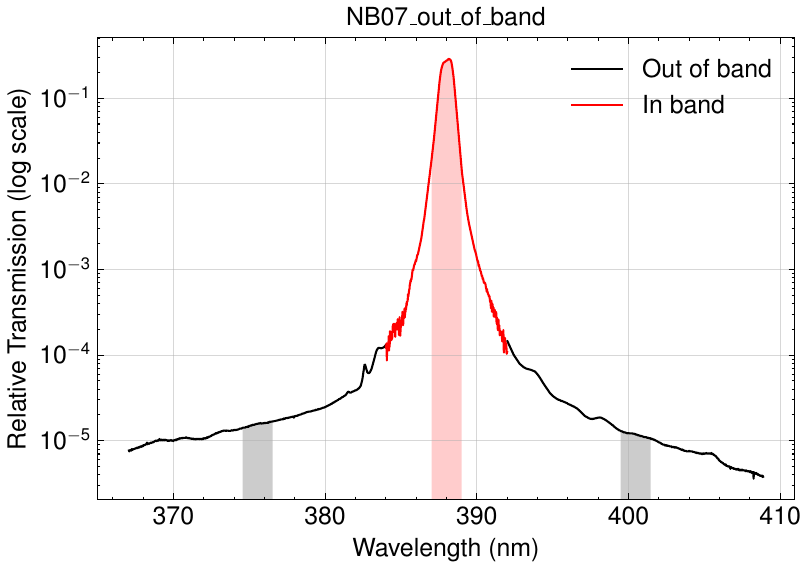}
    \caption{Out-of-band characteristics of \suit narrowband filters. The red and black curves denote separately recorded in-band and out-of-band spectra. The shaded regions correspond to the area under the corresponding spectra used for transmission calculation in Table \ref{tab:oob_table}. For every plot, \textit{Y-axis} represents the relative transmission in \textit{log scale}. The \textit{x-axis} represents the wavelength in \textit{nm}.}
    \label{fig:oob_1}
\end{figure}
\begin{figure}
\centering
    \includegraphics[width = 0.45 \linewidth]{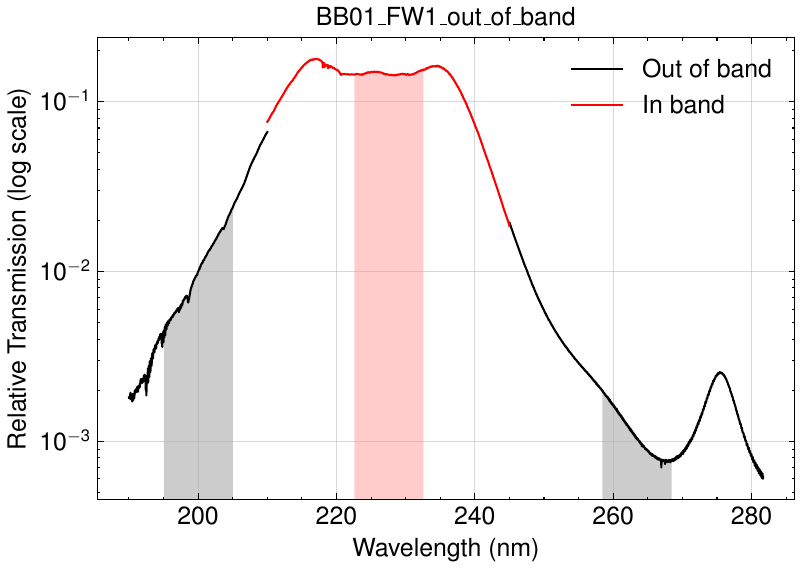}
    \includegraphics[width = 0.45 \linewidth]{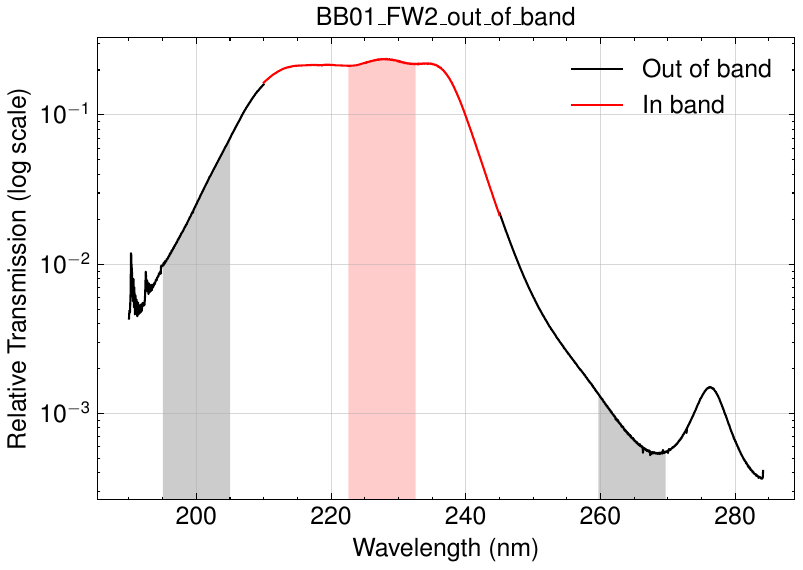}
    \includegraphics[width = 0.45 \linewidth]{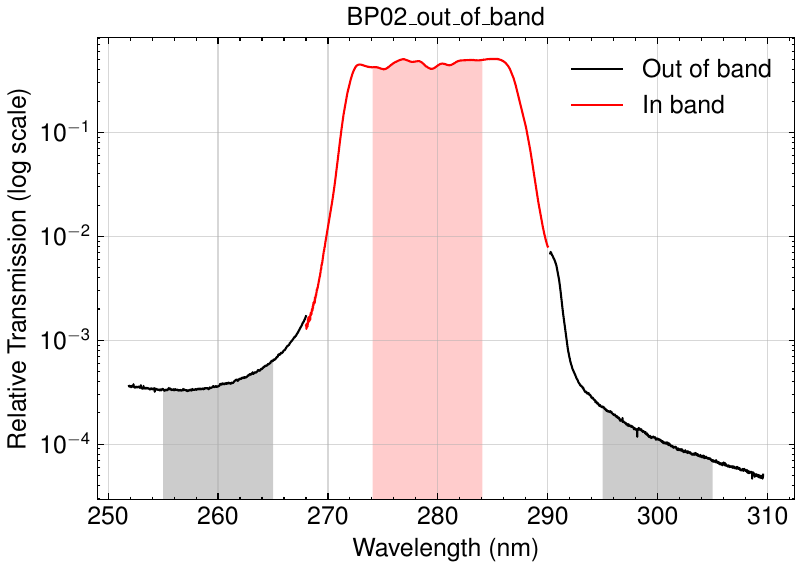}
    \includegraphics[width = 0.45 \linewidth]{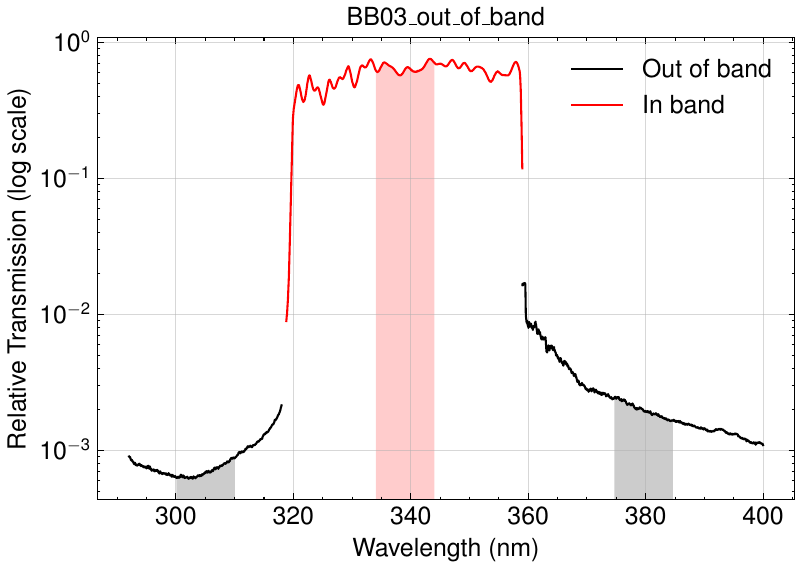}
    \includegraphics[width = 0.45 \linewidth]{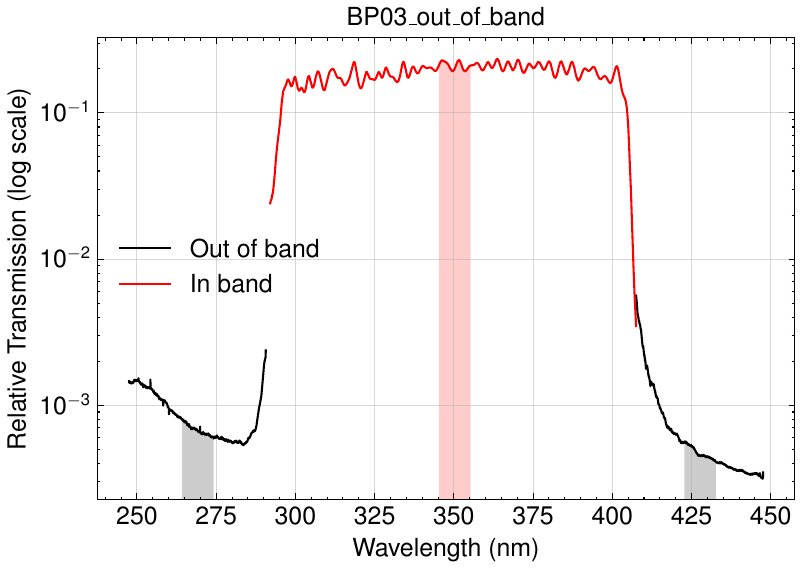}
    \includegraphics[width = 0.45 \linewidth]{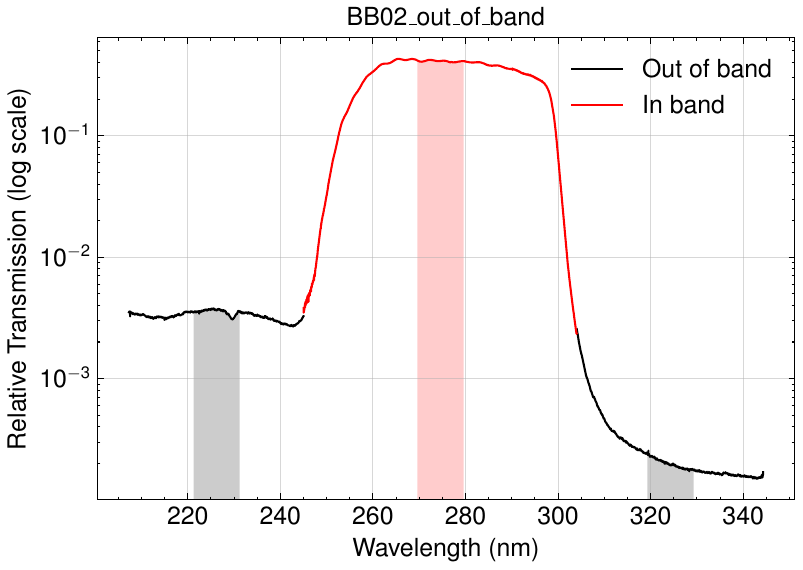}
    \includegraphics[width = 0.45 \linewidth]{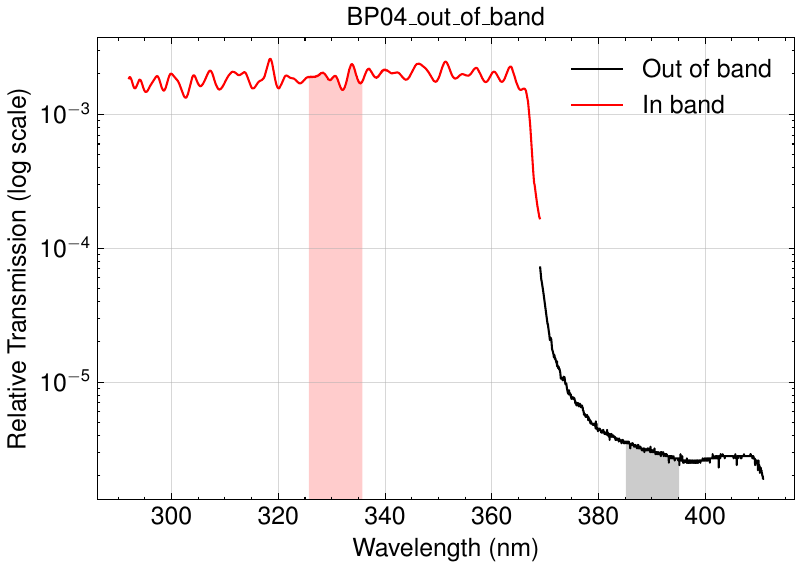}
\caption{Out-of-band characteristics of \suit broadband and bandpass filters. The red and black curves denote separately recorded in-band and out-of-band spectra. The shaded regions correspond to the area under the corresponding spectra used for transmission calculation in Table \ref{tab:oob_table}. For every plot, \textit{Y-axis} represents the relative transmission in \textit{log scale}. The \textit{x-axis} represents the wavelength in \textit{nm}.}
\label{fig:oob_2}
\end{figure}

\begin{figure}
    \centering
    \includegraphics[width = 0.45 \linewidth]{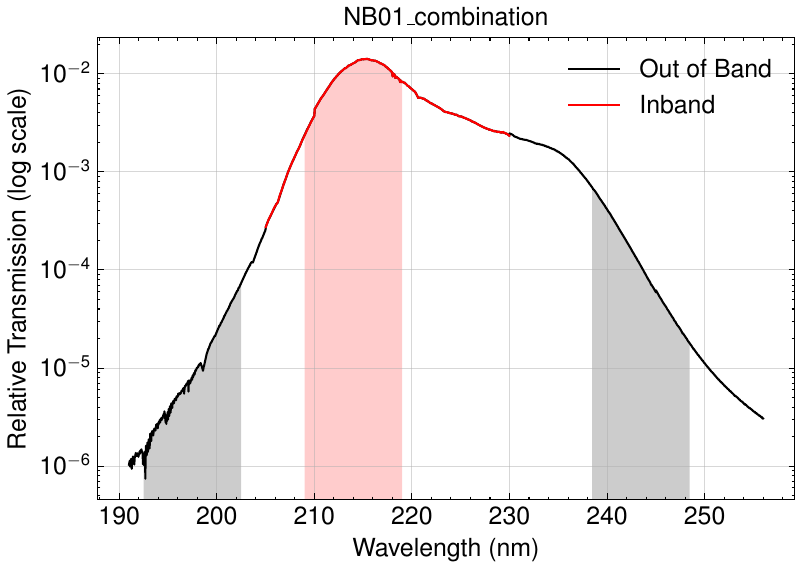}
    \includegraphics[width = 0.45 \linewidth]{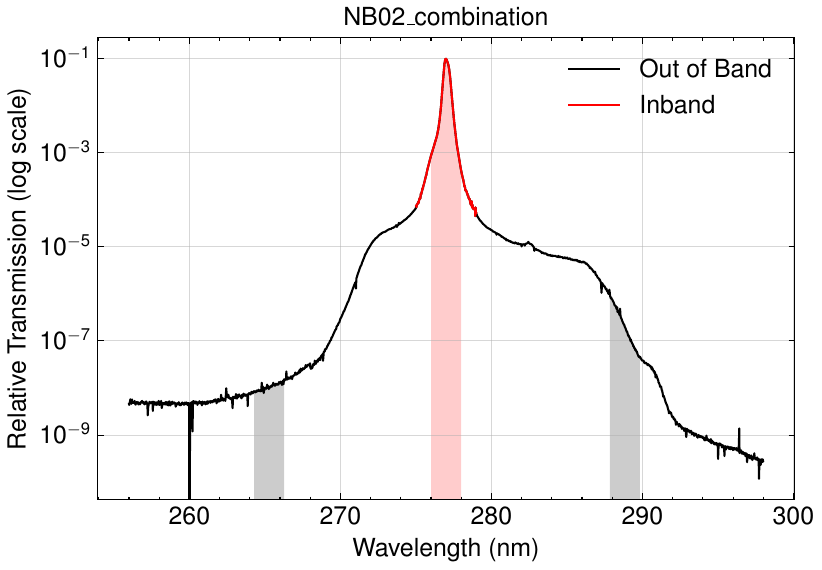}
    \includegraphics[width = 0.45 \linewidth]{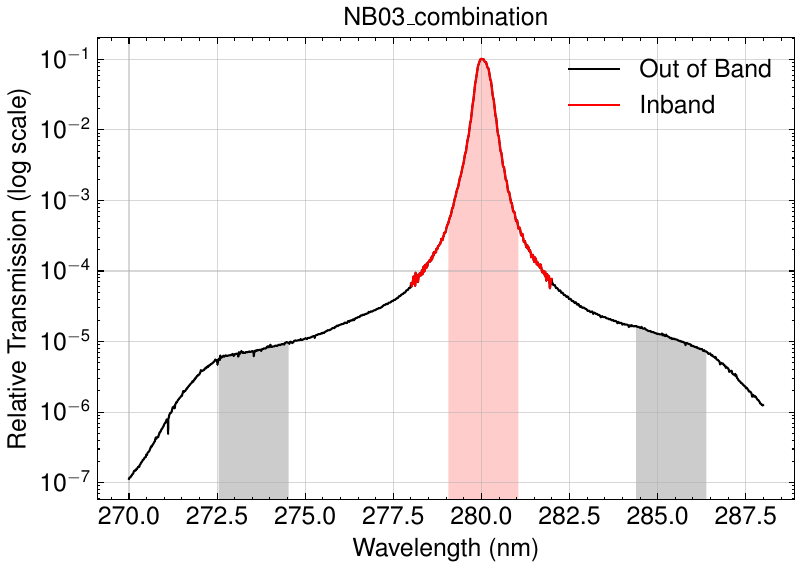}
    \includegraphics[width = 0.45 \linewidth]{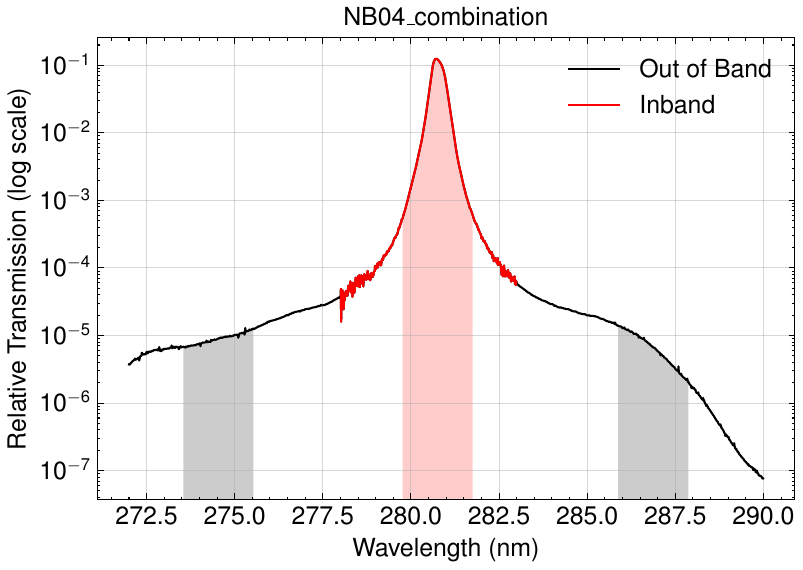}
    \includegraphics[width = 0.45 \linewidth]{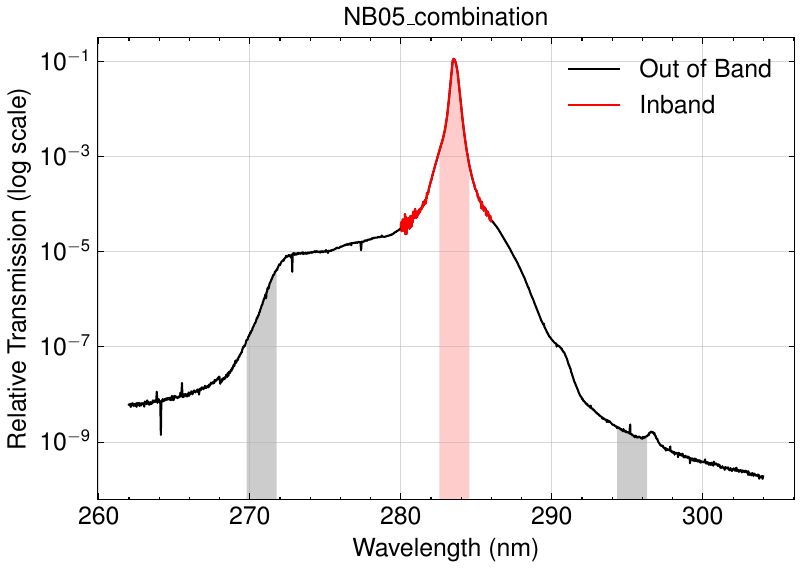}
    \includegraphics[width = 0.45 \linewidth]{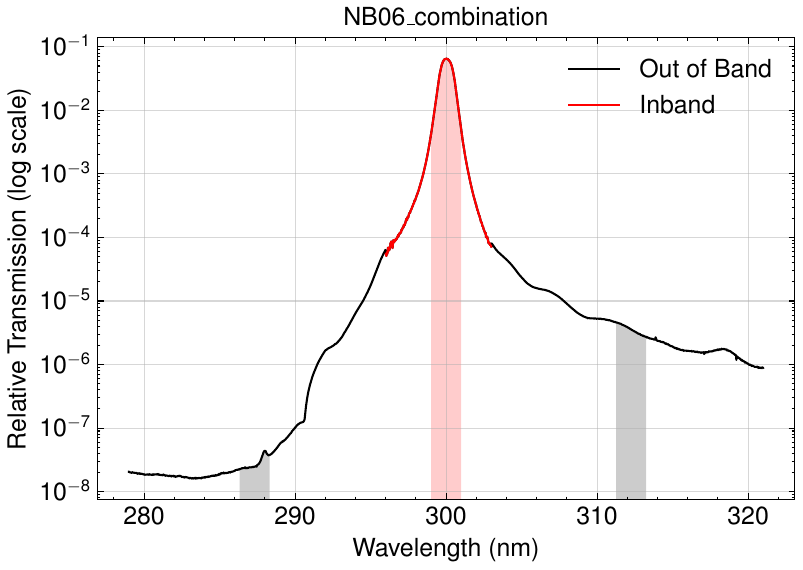}
    \includegraphics[width = 0.45 \linewidth]{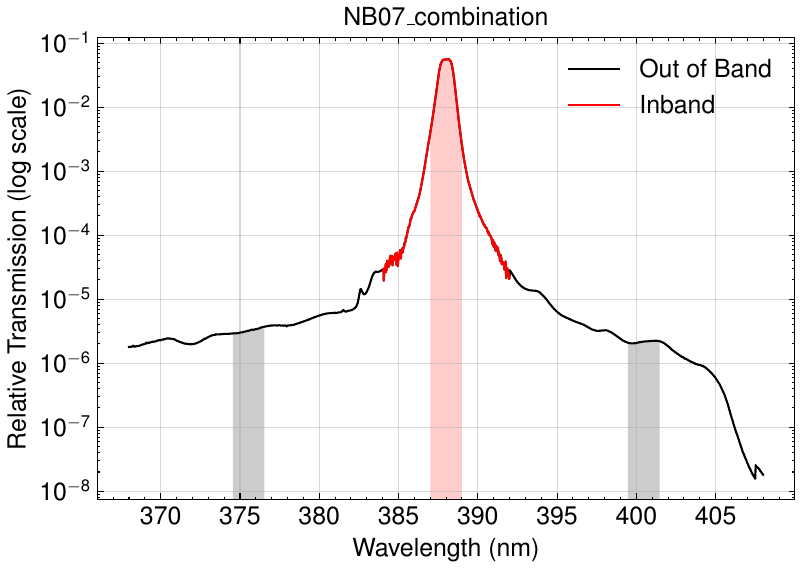}
    \caption{Out-of-band characteristics of \suit narrowband science filter combinations. The red and black curves denote combined in-band and out-of-band spectra. The shaded regions correspond to the area under the corresponding spectra used for transmission calculation in Table \ref{tab:combination_table}. For every plot, \textit{Y-axis} represents the relative transmission in \textit{log scale}. The \textit{x-axis} represents the wavelength in \textit{nm}. \js{The out-of-band profile of the NB08 filter was not recorded due to the risk of contaminating the filter as explained in \S \ref{sec:oob_results}.}}
    \label{fig:combination_1}
\end{figure}
\begin{figure}
\centering
    \includegraphics[width = 0.45 \linewidth]{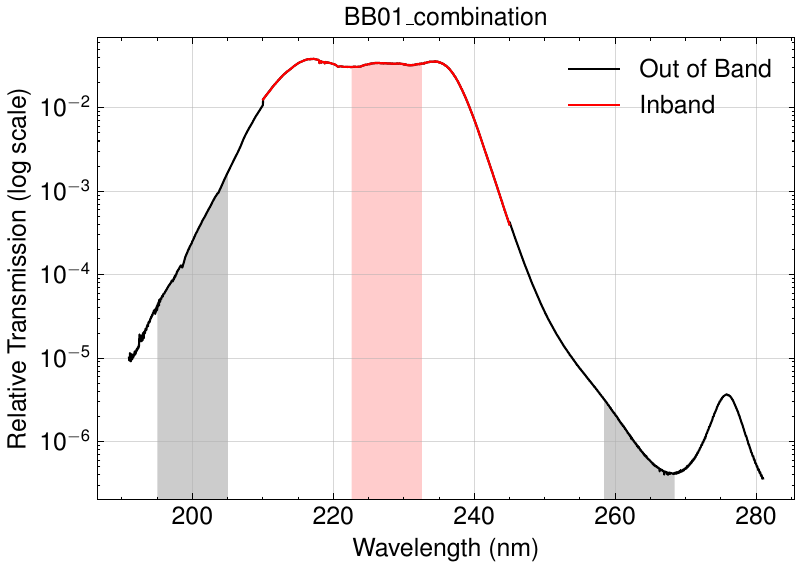}
    \includegraphics[width = 0.45 \linewidth]{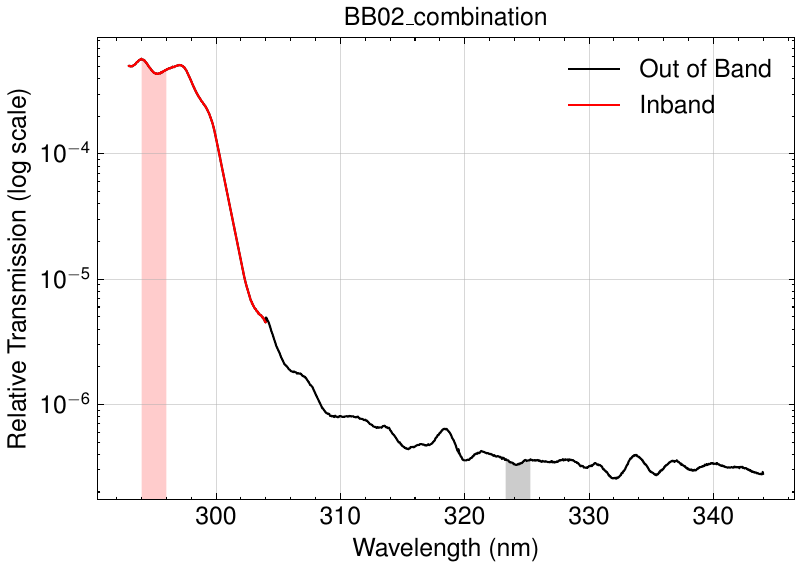}
    \includegraphics[width = 0.45 \linewidth]{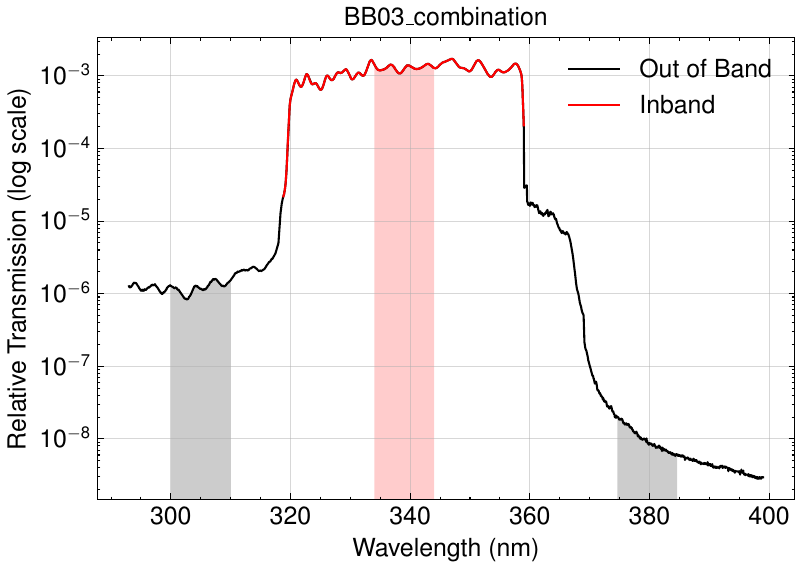}
    \caption{Out of band characteristics of \suit broadband science filter combinations. The red and black curves denote combined in-band and out-of-band spectra. The shaded regions correspond to the area under the corresponding spectra used for transmission calculation in Table \ref{tab:combination_table}. For every plot, \textit{Y-axis} represents the relative transmission in \textit{log scale}. The \textit{x-axis} represents the wavelength in \textit{nm}.}
    \label{fig:combination_2}
\end{figure}

\subsection{Experiment}
Out-of-band transmission refers to the percentage of incident light leaking through the filter at wavelengths beyond the required transmission spectrum. A high out-of-band transmission can lead to reduced contrast and loss of detail. Depending on the filter, SUIT science filters were designed to have an out-of-band transmission of 0.01\% or 0.001\%. 

The spectrum of light transmitted through the filter is recorded from a circular region of diameter 3.86 mm at the center of the filter, as described in \S \ref{sec:ExptSetup}. This is followed by recording the spectrum of the light source passing through all optical components in the setup. The corresponding background frames are recorded and the transmission is calculated using Equation \ref{eqn:tx}.

Significantly high exposure times are required to characterize the out-of-band transmission, as compared to recording the inband or spatial transmission (\S \ref{sec:Spatial}). The exposure time is set such that it is $10^3$ ($10^4$) times the exposure time used to record the spectrum for a filter designed to offer \js{$ < 0.01\%$ (0.001\%)} out-of-band transmission. Following this, the in-band transmission is also measured for the same spatial location of the filter.
The out-of-band transmission is evaluated as a percentage of the inband transmission. A wavelength band of 2 nm (10 nm) for narrowband (broadband) filters is picked around the central transmission wavelength of the filter. The total transmission in this range is integrated within the in-band spectrum. The same is performed for the red and blue out-of-band spectra, which are then presented as a fraction of the in-band transmission.

\subsection{Analysis and Results} \label{sec:oob_results}
\begin{table}
\centering
\begin{tabular}{c c c c}
    \toprule
    \textbf{Filter Name} & \textbf{Blue wing} & \textbf{Red wing} & \textbf{Integration}\\
    & \textbf{(\% Tx)} & \textbf{(\% Tx)} & \textbf{band (nm)}\\
    \midrule
    BB01\footnotemark[1] & $7.66e{+00}$   & $7.86e{-01}$ & 10 \\ 
    BB01\footnotemark[2] & $1.34e{+01}$   & $3.41e{-01}$ & 10 \\ 
    BB02 & $8.62e{-01}$   & $4.85e{-02}$ & 10 \\
    BB03 & $1.09e{-01}$   & $3.08e{-01}$ & 10 \\
    BP02 & $8.65e{-02}$   & $2.65e{-02}$ & 10 \\
    BP03 & $3.23e{-01}$   & $2.26e{-01}$ & 10 \\
    BP04 & NA             & $1.61e{-01}$ & 10 \\
    NB01 & $2.32e{+00}$   & $5.44e{+00}$ & 02 \\
    NB02 & $3.50e{-02}$   & $1.30e{-02}$ & 02 \\
    NB03 & $3.20e{-02}$   & $4.23e{-02}$ & 02 \\
    NB04 & $3.30e{-02}$   & $3.19e{-02}$ & 02 \\
    NB05 & $2.91e{-02}$   & $1.25e{-02}$ & 02 \\
    NB06 & $1.93e{-02}$   & $9.13e{-03}$ & 02 \\
    NB07 & $1.03e{-02}$   & $7.79e{-03}$ & 02 \\
    \botrule
\end{tabular}
\footnotetext[1]{Mounted on filter wheel 1}
\footnotetext[2]{Mounted on filter wheel 2}
\vspace{0.02 \linewidth}
\caption{Percentage of out-of-band transmission on red and blue wings of the filters, with respect to in-band transmission. From left to right, the columns indicate the filter name [col 1], the percentage of transmission in the blue [col 2] and red wing [col 3] with respect to the inband spectrum, and the wavelength band over which the transmission has been integrated [col 4].}
\label{tab:oob_table}
\end{table}

\begin{table}
\centering
\begin{tabular}{c c c c c}
    \toprule
    \textbf{Filter} & \textbf{Filter} & \textbf{Blue Wing} & \textbf{Red Wing} & \textbf{Integration}\\
    \textbf{Wheel 1}& \textbf{Wheel 2} & \textbf{(\% Tx)} & \textbf{(\% Tx)} & \textbf{band (nm)}\\
    \midrule
    BB01 & BB01 & $1.30e{+00}$ & $3.45e{-03}$ & 10 \\ 
    BP04 & BB02 & N/A          & $7.22e{-02}$ & 02 \\
    BP04 & BB03 & $9.45e{-02}$ & $8.14e{-04}$ & 10 \\
    BB01 & NB01 & $1.66e{-01}$ & $1.91e{+00}$ & 02 \\
    NB02 & BP02 & $4.80e{-05}$ & $1.33e{-03}$ & 02 \\
    NB03 & BP02 & $3.12e{-02}$ & $4.73e{-02}$ & 02 \\
    NB04 & BP02 & $3.06e{-02}$ & $2.43e{-02}$ & 02 \\
    NB05 & BP02 & $4.53e{-03}$ & $5.54e{-06}$ & 02 \\
    BP03 & NB06 & $8.38e{-05}$ & $1.04e{-02}$ & 02 \\
    BP03 & NB07 & $1.07e{-02}$ & $7.09e{-03}$ & 02 \\
    \botrule
\end{tabular}
\caption{Percentage of out-of-band transmission on red and blue wings of the filter combinations, with respect to in-band transmission of the filter combination. From left to right, the columns indicate the names of the filters mounted on filter wheels 1 [col 1] and 2 [col 2], making the filter combination, the percentage of transmission in the blue [col 3] and red wing [col 4] with respect to the inband spectrum, and the wavelength band over which the transmission has been integrated [col 5].}
\label{tab:combination_table}
\end{table}

The out-of-band transmission of the filters is measured as mentioned in \S \ref{sec:oob}. The measured spectra are depicted in Figures \ref{fig:oob_1} and \ref{fig:oob_2}. The red curve shows the in-band transmission, while the black curve shows the \js{out-of-band transmission}. The transmissions are integrated within the red and \js{black shaded regions}. The total integrated intensity within a band of 2 nm (or 10 nm) is used to find the fraction of out-of-band transmission with respect to the inband transmission for each filter.

From Table \ref{tab:oob_table}, we notice that all broadband and bandpass filters (except BB01 and NB01, due to the low SNR data for wavelengths below 250 nm) show out-of-band transmissions to be lesser than 1\% with respect to the in-band spectrum on both blue and red wings. For narrow-band filters, the out-of-band transmission is below 0.1\% with respect to the in-band transmission. This suggests that the in-band transmission of the filters is significantly stronger than the out-of-band transmission. 


In the case of the BP04 filter, the peak in-band transmission is of the order of 0.1. Moreover, the xenon lamp intensity and spectrograph sensitivity below 250 nm wavelengths is very low because of atmospheric attenuation of UV. This leads to extremely long exposure times (~1200 s) for accumulating enough counts on the blue side of the spectrum. Therefore, this data was not recorded to prevent the filter from being exposed for a long duration. 
The same is true for the out-of-band characterization of filters delivering low throughput. Therefore, the out-of-band profile of the NB08 filter was not recorded due to the risk of contamination.

Since the \suit bandpasses operate with two filters in tandem, a similar bandpass analysis was done by combining the transmissions of both filters for each filter combination, which is depicted in Figures \ref{fig:combination_1} and \ref{fig:combination_2}. The same area of integration, as used for Table \ref{tab:oob_table}, was used to evaluate the relative out-of-band transmission with respect to the inband transmission for the filter combinations. The results are tabulated in Table \ref{tab:combination_table}. We note that the out-of-band transmission for most filters \js{is} several orders of magnitude lower than 1\%. This is except for filters BB01 and NB01, which operate at wavelengths below 250 nm, thus having low SNR due to high atmospheric attenuation of UV and low intensity of incident light from the Xenon lamp at these wavelengths. \js{The high out-of-band to the inband ratio for the NB01 and BB01 filter combinations is due to the filter manufacturing limitations at these short wavelengths, below 250 nm.}
The combined transmission for BB02 could not be produced as its combination filter, BP04, lacks data at the blue wing.

\section{Tilt Characterization}\label{sec:tilt}
\begin{figure}
\centering
    \includegraphics[width = 0.45 \linewidth]{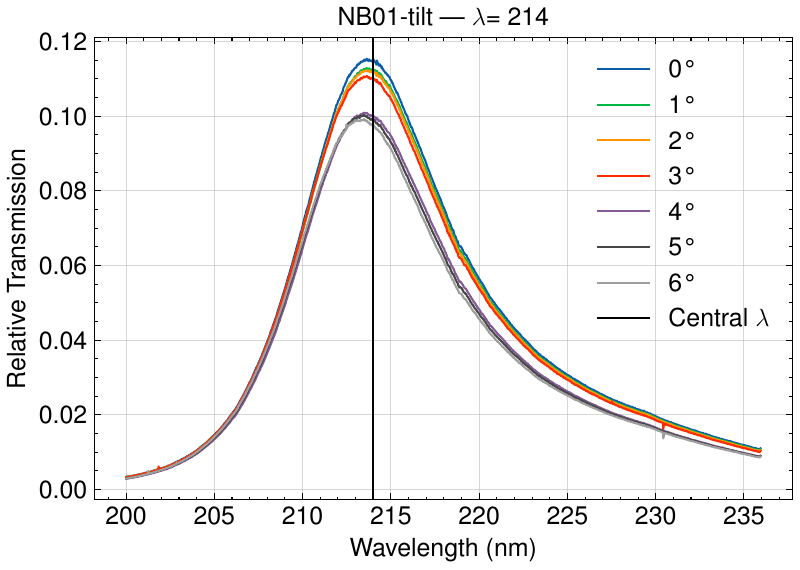}
    \includegraphics[width = 0.45 \linewidth]{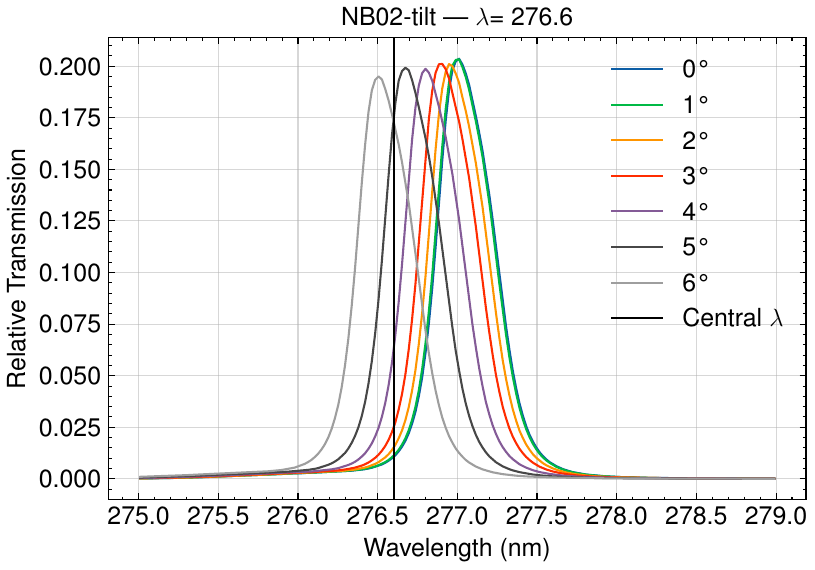}
    \includegraphics[width = 0.45 \linewidth]{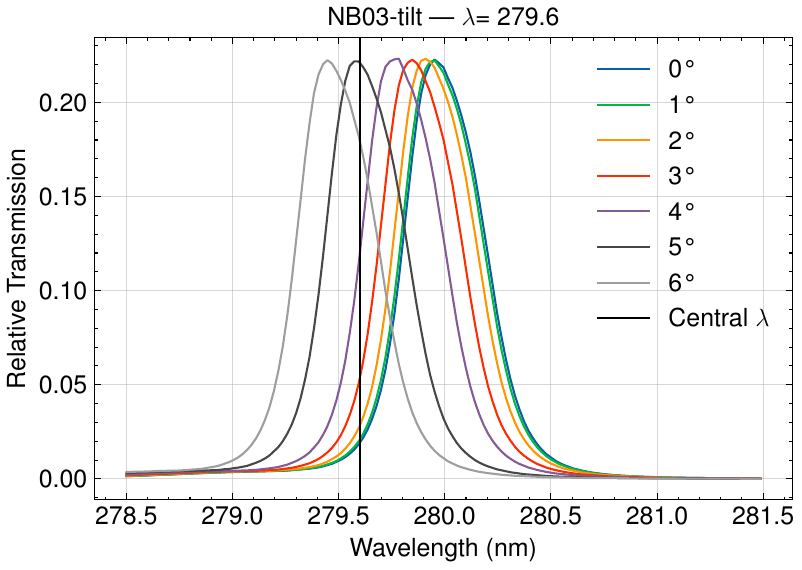}
    \includegraphics[width = 0.45 \linewidth]{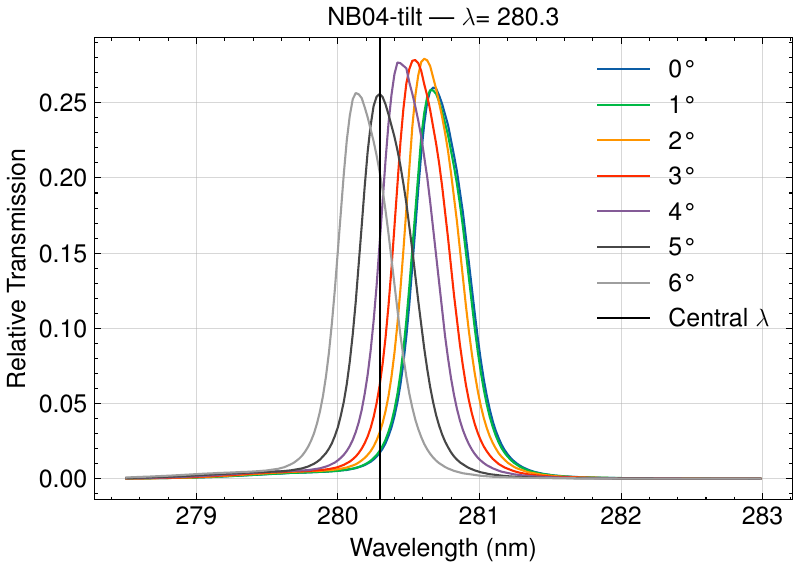}
    \includegraphics[width = 0.45 \linewidth]{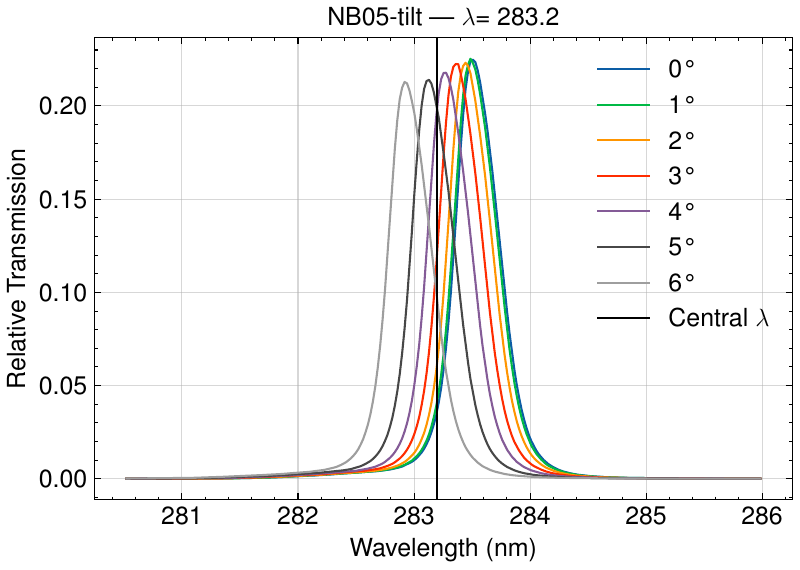}
    \includegraphics[width = 0.45 \linewidth]{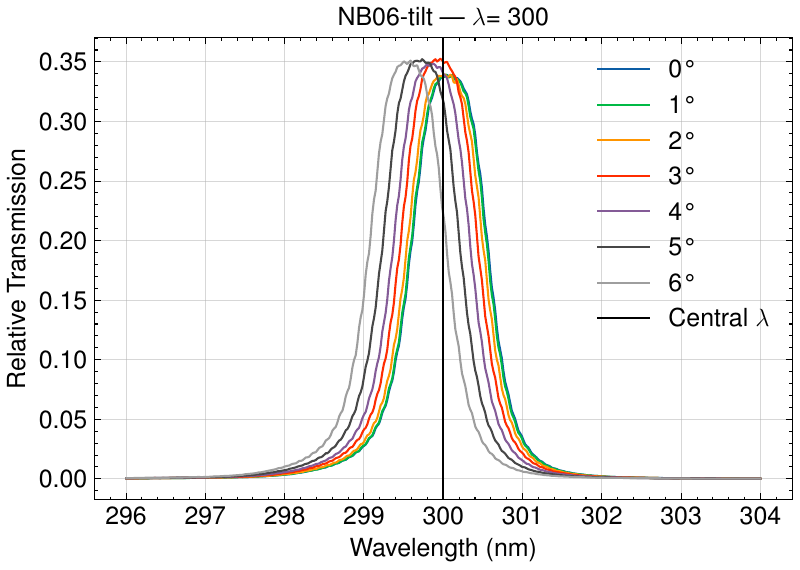}
    \includegraphics[width = 0.45 \linewidth]{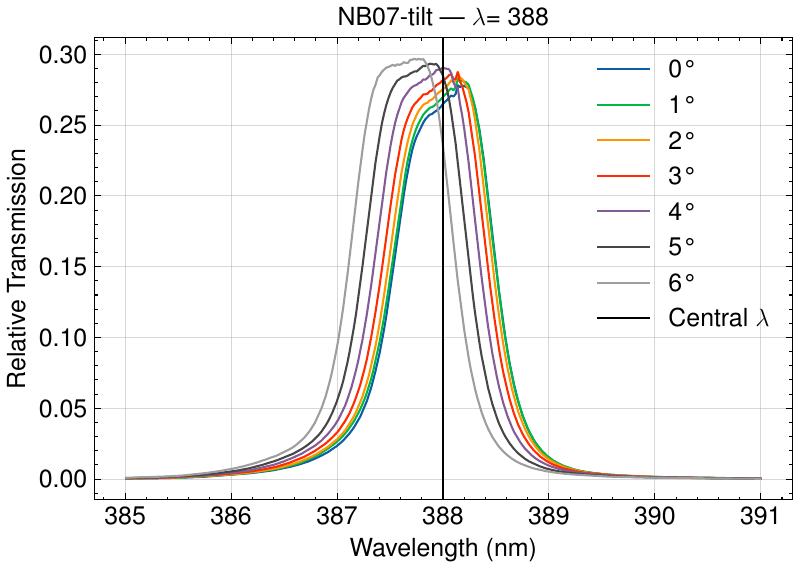}
    \caption{Transmission profiles for various tilt angles for \suit narrowband filters. The various colored curves denote the transmission spectrum at various tilt angles. The black vertical line marks the central wavelengths of scientific interest. The tilt angle is chosen based on the curve that best overlaps with the black vertical line, which is also tabulated in Table \ref{fig:tilt}. The target wavelength of observation is indicated at the top of every plot.}
    \label{fig:tilt}
 \end{figure}
\subsection{Experiment}
Light incident on two parallel dichroic filters can undergo multiple reflections from each other, leading to ghost images. To mitigate this, the two filters delivering one bandpass are tilted away from each other to minimize ghost reflections between the filters and subsequent optical components.

The transmission of dichroic filters shifts towards the blue side of the spectrum with the increase in the angle of incidence. \suit science filters are fabricated with the central wavelengths deliberately shifted to a wavelength longer than required, which can then be tilted to get the right bandpass and minimize ghost reflections. Optimizing the tilts and ensuring maximum throughput at the required wavelength is necessary to observe specific line emissions of the Sun. 

A rotation stage is placed in the beam path and collimated light is made incident on the filter as described in \S \ref{sec:ExptSetup}. A filter is tilted in 1-degree steps on either side, and the transmission is recorded. The desired wavelength to be observed through these filters is plotted, and the transmission profile peaking closest to the desired wavelength is chosen for mounting on the filter wheel. The most appropriate mounting ring is chosen to introduce the optimal tilt angle in each filter accordingly. The transmission of broadband and bandpass filters is typically in the order of 40 nm. The shift in bandpass introduced with tilt is of the order of 0.1 nm, which is negligible to the transmission spectrum for these filters. These filters are mounted with high tilt angles ($>4 {\arcdeg}$) to minimize ghost reflections. The derived optimal tilt angles and the tilt finally applied to the filter while mounting on the filter wheel are tabulated in Table \ref{table:tilt}.

\subsection{Analysis and Results}

\begin{table}
    \centering
    \begin{tabular}{c c c}
        \toprule
        \textbf{Filter} & \textbf{Optimal Tilt} & \textbf{Applied Tilt}\\ 
         & \textbf{(deg)} & \textbf{(deg)}\\ 
        \midrule
        NB01 & 6 & 6 \\ 
        NB02 & 6 & 6 \\ 
        NB03 & 5 & 5 \\ 
        NB04 & 5 & 5 \\ 
        NB05 & 5 & 5 \\ 
        NB06 & 3 & 4 \\ 
        NB07 & 4 & 4 \\ 
        NB08\footnotemark[1] & 0 & 0 \\ 
        NB08\footnotemark[2] & 0 & 0 \\
        BB01 & Any & 4 \\
        BB01 & Any & 4 \\
        BB02 & Any & 6 \\
        BB03 & Any & 4 \\
        BP02 & Any & 6 \\
        BP03 & Any & 5 \\
        BP04 & Any & 4 \\
        \botrule
    \end{tabular}
    \footnotetext[1]{Mounted on filter wheel 1} 
    \footnotetext[2]{Mounted on filter wheel 2}
    \vspace{0.02\linewidth}
	\caption{Optimal tilt angles applied to \suit filters upon mounting. The columns from left to right indicate the filter name [col 1], the optimal angle of tilt as derived from Figure:\ref{fig:tilt} [col 2] and the applied angle of tilt [col 3].}
	\label{table:tilt}
\end{table}
The transmission spectra at successive tilt angles are plotted in Figure \ref{fig:tilt}. The black vertical line in each plot denotes the desired target wavelength for each filter. It is noted that NB01 has a wide bandpass. Therefore, the relative shift in central wavelength due to tilt is relatively small. A high tilt angle \js{of $6^\circ$ is implemented on NB01} to reduce the possibility of ghost image formation. The optimal tilts for the other filters are estimated from the desired central wavelengths and tabulated in Table \ref{table:tilt}. The optimal tilt for NB06 was seen to be 3{\arcdeg}. But due to mechanical constraints, 4{\arcdeg}  tilt is applied to the filter. However, science goals are not affected as this band observes a continuum region in the solar spectrum. The tilt for the NB08 filter, meant to observe the Ca-h line at 396.85 nm, is kept at 0{\arcdeg}. This filter has a bandpass of 0.1 nm, and any spurious tilts to this filter can lead to a high relative shift in wavelength, causing the Ca-h line to fall out of the filter's in-band spectrum. This compromise comes with the increased possibility of ghost reflections in the images.
	
\section{Mounting of science filters}\label{sec:mounting}

\begin{figure}
    \centering
    \includegraphics[width=0.9\linewidth]{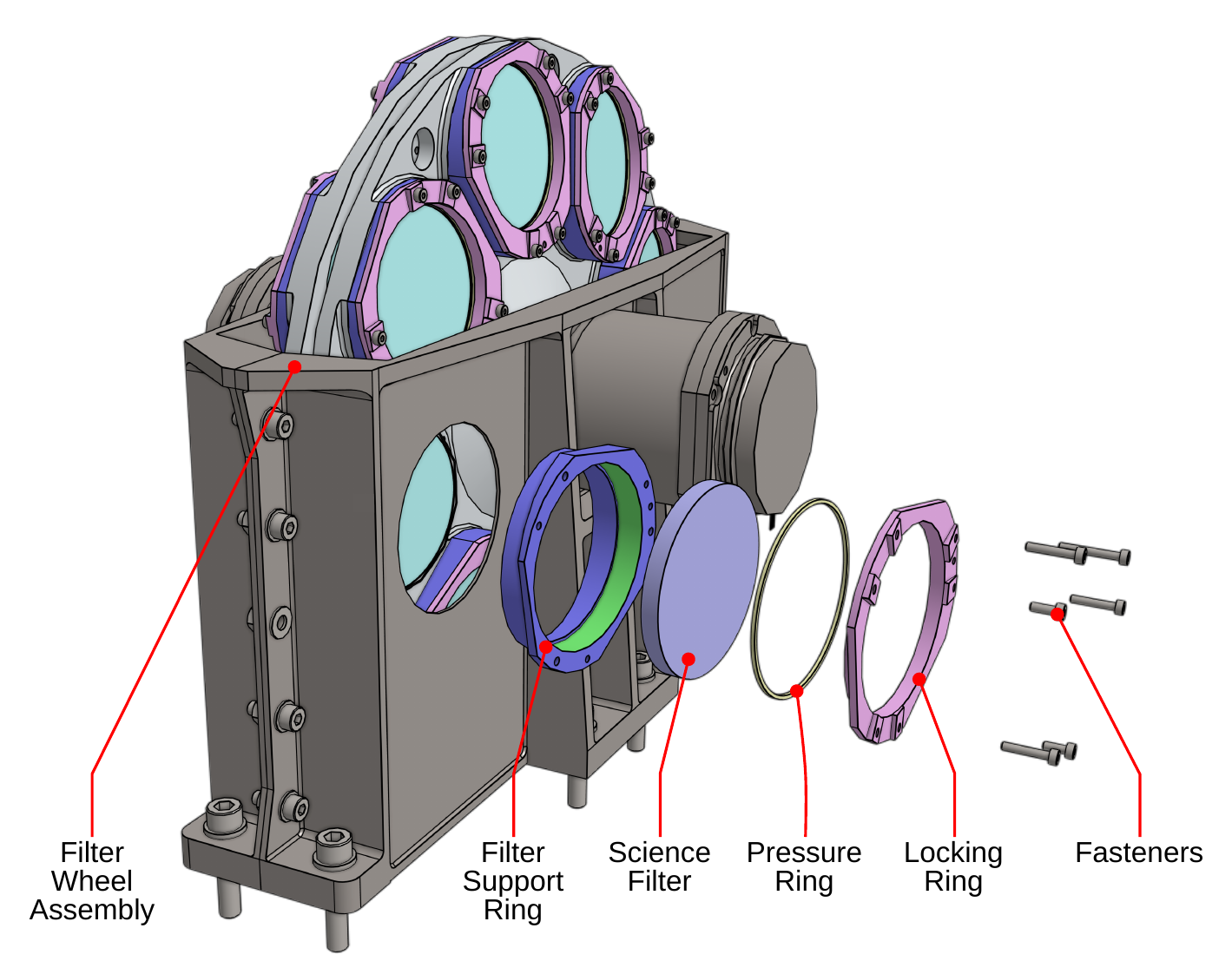}
    \caption{Exploded view of the science filter mounting system on the \suit filter wheel assembly. \suit has 16 filters mounted on two independently rotating filter wheels with 8 slots each. Each filter is mounted on the filter wheel with a support ring, a pressure ring, and a locking ring, held onto the filter wheel with 6 fasteners.}
    \label{fig:exploded_diag}
\end{figure}
The \suit science filter assembly uses two independently rotating filter wheels, positioning appropriate filter combinations in the beam path to create the desired bandpasses. The position of the filters on each wheel is determined to minimize the movement of the filter wheel motors during the imaging sequences.

Selected filters of each type are mounted at their predetermined position with the help of a support ring- that acts like a pocket and houses the filter, a pressure ring- that maintains uniform pressure on the filter upon mounting, and a lock ring- that locks the filter in place restricting any form of lateral or rotational movement. This setup is attached to the filter wheel with six fasteners as depicted in Figure \ref{fig:exploded_diag}.

The science filters are made of two glass elements mounted on a black anodized aluminum ring. The filters have a demarcation on the ring denoting their name, serial number, and the direction facing the incident light. The substrate for the filter is fused silica for its high transmission in the UV band. One layer of the science filter helps cut down excess sunlight; the other is a dichroic layer, which is necessary to pass the appropriate spectral band of sunlight. The filters are mounted such that light is first incident on the intensity reduction layer, followed by the dichroic layer.

The filters were mounted in a Class 100 clean room to prevent particulate or molecular contaminants from deposition on the glass. For every filter, the marker for the direction of incident light on the rim is set at a specific clocking orientation. This enables us to understand the spatial nature of transmission through the filter based on the spatial transmission data measured in \S \ref{sec:Spatial}.

The filter and mounting rings are locked on the filter wheel with six symmetrically placed titanium alloy M2 fasteners driven into self-locking stainless steel thread inserts in the filter wheel. While mounting, torque is applied on diametrically opposite bolts in steps of 10 N-cm beyond the running torque to prevent any stress or tilt on the filters. All the bolts are torqued to the Indian Space Research Organization (ISRO) recommended 40 N-cm, required for safe spaceflight qualification. The bolt heads were potted with 3M EC2216 adhesive for added rigidity.

\section{Summary and Discussion}\label{sec:conclusion}
In this paper, we have presented methodology and results for testing the science filters mounted on \suit. We present the qualification test results and perform optical tests to evaluate filter performance. We prepared an optical setup where we collimated light from a xenon arc lamp and illuminated a portion of the filter. The light transmitted from the filter was focused into the slit of a spectrometer, and the transmission spectrum was recorded. 

We characterized the spatial variation in transmission, the variation in transmission due to the angle of tilt, and the out-of-band transmission. 
The spatial variation of transmission is satisfactory for all filters. The maximum relative fluctuation in central wavelength is $2.09e-2~nm$ for a bandpass with FWHM of $0.13~nm$ for the NB08 filter. We have also characterized the transmission variation for the narrowband filters at different tilt angles to find the most appropriate mounting angle. The tilt is introduced to minimize ghost reflections while care is taken to get the right bandpass for specific filters.
The out-of-band transmission is measured with respect to the in-band transmission for the filters. The out-of-band to in-band ratio for two filter combinations is in the \js{$>1\%$ domain}, and that for three filters is in the \js{$>0.1\%$ domain}, with the rest being in the order of 0.01\%. \js{The reason for the high out-of-band to the inband ratio for the NB01 and BB01 filter combinations is due to the filter manufacturing limitations at these short wavelengths.} 

\js{The results obtained here demonstrate that the science filters on \suit meet the necessary performance criteria, ensuring good image quality and reliable photometric measurements across the telescope's field of view. Moreover, the performance of various filters in the NUV and mid-UV bands may be used as a benchmark for future missions for solar and stellar observations in this wavelength band.}


\bmhead{Acknowledgements}

\js{We thank the reviewer for the constructive comments and suggestions}. \suit is built by a consortium led by the Inter-University Centre for Astronomy and Astrophysics (IUCAA), Pune, and supported by ISRO as part of the Aditya-L1 mission. The consortium consists of SAG/URSC, MAHE, CESSI-IISER Kolkata (MoE), IIA, MPS, USO/PRL, and Tezpur University. Aditya-L1 is an observatory class mission which is funded and operated by the Indian Space Research Organization. The mission was conceived and realised with the help from all ISRO Centres and payloads were realised by the payload PI Institutes in close collaboration with ISRO and many other national institutes - Indian Institute of Astrophysics (IIA); Inter-University Centre of Astronomy and Astrophysics (IUCAA); Laboratory for Electro-optics System (LEOS) of ISRO; Physical Research Laboratory (PRL); U R Rao Satellite Centre of ISRO; Vikram Sarabhai Space Centre (VSSC) of ISRO. This research used PYTHON packages NumPy \cite{numpy}, Matplotlib \cite{matpltolib}, SciPy \cite{scipy}, and SciencePlots \cite{SciencePlots}. The 3D illustrations were made in Onshape \cite{onshape}. The science filters for \suit were manufactured by Materion Corp. and were acquired through the \suit Project.
\bmhead{Data Availability}
The processed data and results are presented in the tables and figures of the paper, which are available upon request.

\bibliography{bib_db}
\end{document}